%% file: main.tex
\documentclass[sigconf,9pt, table, xcdraw,authorversion,nonacm]{acmart}

\acmConference[Sigcomm]{ACM Sigcomm}{September 8--11,
  2025}{Coimbra, PT}

\usepackage{algorithmic}
\usepackage{graphicx}
\usepackage{textcomp}
\usepackage{algorithm}
\usepackage{subcaption}
\usepackage{booktabs}
\usepackage{multirow}
\usepackage{tabularx} 
\usepackage{comment}
\usepackage{xspace}
\usepackage{makecell}
\usepackage{enumitem}
\usepackage{tikz}
\usepackage{url}
\usepackage{newfloat}
\usepackage{listings}
\usepackage{soul}
\usepackage[normalem]{ulem}

\begin{document}

\title[Debunking Representation Learning]{The Sweet Danger of Sugar: Debunking Representation Learning for Encrypted Traffic Classification}

\author{Yuqi Zhao}
\affiliation{%
  \institution{Politecnico di Torino}
  \city{Torino}
  \country{Italy}}
\email{yuqi.zhao@polito.it}

\author{Giovanni Dettori}
\affiliation{%
  \institution{Politecnico di Torino}
  \city{Torino}
  \country{Italy}}
\email{giovanni.dettori@polito.it}

\author{Matteo Boffa}
\affiliation{%
  \institution{Politecnico di Torino}
  \city{Torino}
  \country{Italy}
  }
\email{matteo.boffa@polito.it}

\author{Luca Vassio}
\affiliation{%
  \institution{Politecnico di Torino}
  \city{Torino}
  \country{Italy}}
\email{luca.vassio@polito.it}

\author{Marco Mellia}
\affiliation{%
  \institution{Politecnico di Torino}
  \city{Torino}
  \country{Italy}}
\email{marco.mellia@polito.it}

\renewcommand{\shortauthors}{Zhao et al.}
\newcommand{\ie}{\mbox{i.e.,}\xspace}
\newcommand{\eg}{\mbox{e.g.,}\xspace}
\newcommand{\tool}{\mbox{\textit{Pcap-Encoder}}\xspace}
\newcommand{\VPNBin}{\mbox{VPN-Binary}\xspace}
\newcommand{\VPNSer}{\mbox{VPN-Service}\xspace}
\newcommand{\VPNApp}{\mbox{VPN-App}\xspace}
\newcommand{\USTCBin}{\mbox{Malware}\xspace}
\newcommand{\USTCApp}{\mbox{App-19}\xspace}
\newcommand{\CSTNETTLS}{\mbox{TLS-120}\xspace}

\begin{abstract}
\input{sections/00_Abstract/abstract}
\end{abstract}



\maketitle

\section{Introduction}\label{sec:intro}
\input{sections/01_Introduction/intro}

\section{Representation Learning: Core Principles}\label{sec:replearn}
\input{sections/02_Core_Principles/core_principles}

\section{Representation Learning: Network Traffic}\label{sec:net_repr}
\input{sections/03_Network_Repr_Learning/01_intro_repr_for_traff}
\input{sections/03_Network_Repr_Learning/02_generic_howto_repr_4_traff}
\input{sections/pcapencoder}

\section{Benchmark for Network Traffic Classification}\label{sec:training_pipeline}
\input{sections/04_Pipeline/pipeline}

\section{Experimental setup}
\label{sec:setup}
\input{sections/05_Results/results}

\section{Related Works}\label{sec:related_work}
\input{sections/07_Related_Work/related_work}

\section{Conclusion}
\label{sec:conclu}
\input{sections/06_Conclusions/conclusions}

\begin{acks}
This work has been funded by Huawei Technologies France under the project ``AISN – AI Secured Networks: Novel Approaches for Concept-Constrained multi-modal Learning for generalizable task-specific Language Models'', which proved fundamental to the genesis of the work. 
Yuqi Zhao has been supported by the China Scholarship Council (Grant No. 202306470001). Matteo Boffa has been supported by the AI4CTI FISA project \#FISA-2023-00168 funded by the Italian Ministry of University and Research (MUR).
Marco Mellia has been supported by the project SERICS (PE00000014) under the MUR National Recovery and Resilience Plan funded by the European Union - NextGenerationEU. 
Computational resources were provided by HPC@POLITO (\url{https://hpc.polito.it}).
\end{acks}

\bibliographystyle{ACM-Reference-Format}
\bibliography{main.bib}
\appendix
\section{Appendix}
\input{sections/Appendix_tool}

\end{document}

%% file: sections/00_Abstract/abstract.tex
Recently we have witnessed the explosion of proposals that, inspired by Language Models like BERT, exploit Representation Learning models to create traffic representations. All of them promise astonishing performance in encrypted traffic classification (up to 98\% accuracy). 
In this paper, with a networking expert mindset, we critically reassess their performance.
Through extensive analysis, we demonstrate that the reported successes are heavily influenced by data preparation problems, which allow these models to find easy \textit{shortcuts} -- spurious correlation between features and labels -- during fine-tuning that unrealistically boost their performance.
When such shortcuts are not present -- as in real scenarios -- these models perform poorly.
We also introduce \tool, an LM-based representation learning model that we specifically design to extract features from protocol headers. \tool appears to be the only model that provides an instrumental representation for traffic classification. Yet, its complexity questions its applicability in practical settings.
Our findings reveal flaws in dataset preparation and model training, calling for a better and more conscious test design. We propose a correct evaluation methodology and stress the need for rigorous benchmarking.

%% file: sections/01_Introduction/intro.tex
We are witnessing the success of Artificial Intelligence (AI), with Deep Neural Networks (DNN), Large Language Models (LLM) and multimodal models empowering applications in several fields. Pre-training using self-supervised tasks is the driving factor behind this `AI Boom'~\cite{wiredGooglesGemini}. 
Self-supervision trains models to solve pretext tasks -- such as next-word (for text) or patch (for images) predictions~\cite{devlin2018bert,he2022masked} -- on humongous unlabelled datasets.
Through this first \textit{representation learning}~\cite{cnbcthirstyGenerative} phase, pre-trained models learn to broadly master the nuances of the input data, learning how to turn texts or images into meaningful \emph{embeddings}, \ie compact yet highly informative numerical representations of the data. Later, such embeddings become useful to solve real-world tasks, called \textit{downstream tasks}.
This two-stage approach has proven extremely successful when solving a specific task given a few or even no examples (few-shot~\cite{brown2020language} or zero-shot learning~\cite{longpre2023flan}).

\begin{figure}[!t]
    \centering
    \includegraphics[width=\columnwidth]{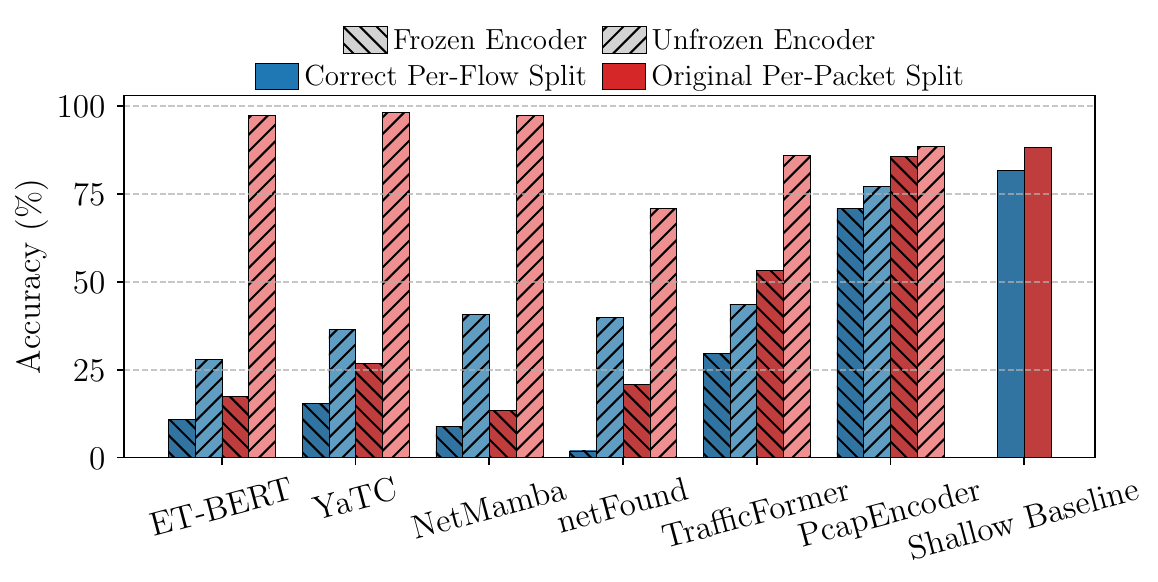}
    \caption{Accuracy of classifiers evaluated (TLS-120 dataset, packet classification task). Models' performance collapses when properly tested. \tool is the only model that maintains good performance. However, a simple shallow baseline surpass all representation learning-based methods.}
    \label{fig:tls_results}
\end{figure}

The allure of AI has captivated many, leading to a surge in the adoption of AI-based solutions to address network traffic classification~\cite{finsterbusch2013survey,nguyen2008survey,rezaei_deep_2019,pacheco_towards_2019,papadogiannaki2021survey}, where encryption makes deep packet inspection ineffective~\cite{naylor2014cost,trevisan2020five}.
Like `Bidirectional Encoder Representations from Transformers' (BERT)~\cite{devlin2018bert} learns to represent text, `Encrypted Traffic BERT' (ET-BERT)~\cite{lin2022etbert} learns to represent packets, even if encrypted.

Most prior work adopts well-established AI pre-training pipelines and architectures.
When fine-tuned for traffic classification, they reach up to 98\% accuracy on VPN or TLS-encrypted traces. Some of the authors -- quite controversially -- claim that pre-training 
allows the model to extract patterns from the encrypted payloads~\cite{lin2022etbert}. 

As network experts, we must assess whether these proposals live up to their astonishing performance. What information can these approaches extract in an everything-encrypted setup? Do these models extract a meaningful traffic representation, or do they simply exploit \textit{shortcuts} -- spurious correlations between features and labels~\cite{geirhos2020shortcut} -- to predict in the downstream task?

This paper presents a systematic critical view of the adoption of representation learning for traffic classification. 
To ensure a fair comparison between different approaches, we define benchmarks based on open datasets, where these models face increasingly complex tasks, up to identifying traffic from 120 websites from TLS traces.
Following ML principles and with a network expert mindset, we introduce a pipeline to properly assess the models' performance.
We pay particular attention to possible pitfalls during cleaning, splitting, sampling and training.
In this common playground, we compare the performance of
state-of-the-art representation learning models. 
Among contenders, we include \tool, our new proposal based on the Text-to-Text Transfer Transformer (T5)~\cite{raffel2020exploring} that we train specifically to extract the \textit{format} and \textit{semantics} of packet headers and to ignore any (encrypted) payload.

We highlight pitfalls that previous works underestimated and that, with promising sweet results, greatly poison the evaluation. Figure~\ref{fig:tls_results} summarises our findings:
    
    \vspace{0.1cm}$\bullet$ The \textit{per-packet split} policy adopted by most of the previous work does not properly separate training samples from testing ones. This creates a huge data leak that creates shortcuts these complex models immediately exploit. Using \textit{per-flow split}, \ie including all packets from the same flow either in the training or testing set as in a real-world setup,
    suffices to remove the most prominent shortcuts. The classification accuracy barely reaches 40\%. 
    
    \vspace{0.1cm}$\bullet$  In the downstream tasks, all previous works train the entire classification architecture (\textit{unfrozen} encoder) with hundreds of thousands of samples. This `destroys' the pre-trained information and basically re-trains from scratch the entire model. When the embedders are frozen (\textit{frozen} encoder), the accuracy of the models drops below 30\%.
    This questions the representation learning abilities and confirms the intuition that training a model to learn patterns from the encrypted payload makes little sense. 

    \vspace{0.1cm}$\bullet$ \tool is the only model that provides a meaningful and robust representation. By design, it exploits packet headers' information and ignores payload. However, the shallow baseline performs on par or better, with much less complexity. This calls into question representation learning at large for its practical applicability.

In summary, the pre-trained models presented in the literature fall short of producing an informative representation for traffic classification. 
The results we present call for a more cautious and critical view of representation learning in traffic analysis. When results appear surprisingly strong -- especially where domain knowledge suggests limited learnability (e.g., encrypted traffic) -- it is essential to examine why the model performs well. 

We recommend the following practices:

\begin{itemize}
    \item \textbf{Control for shortcut learning} — Be aware that deep models are particularly prone to leveraging unintended patterns instead of genuinely learning the task of interest.
    \item \textbf{Verify data integrity} — Ensure that the dataset is free from leakage, artefacts, or spurious correlations that the ML model would immediately exploit as shortcuts.
    \item \textbf{Stress representation learning capabilities} — Assess whether the model is truly learning useful representations, i.e., freeze the encoder during downstream training.
    \item \textbf{Consider cost-benefit trade-offs} — Compare the proposed approach against simple baselines to determine whether the added complexity of deep learning models is justified.
\end{itemize}

We believe our lesson extends to all AI-based solutions for computer networks. To this end, we provide the code, benchmark datasets and methodology to the community to establish a shared environment for development and testing\footnote{\url{https://github.com/SmartData-Polito/Debunk_Traffic_Representation}}.

%% file: sections/02_Core_Principles/core_principles.tex
We introduce the fundamentals and intuitions of representation learning for readers who are not experts in the field.

\vspace{0.1cm}\textbf{Representation learning.}
Machine learning models, as computational systems, inherently work with numerical vectors. The primary goal of \textit{representation learning} is to learn `meaningful' mappings that encode the real properties of an input into a numerical space called the \textit{embedding space}. The component in charge of learning this mapping is named \textit{Encoder} which is \textit{pre-trained} using pretext and self-supervised tasks. 
An embedding space is meaningful when it respects and captures the initial data properties. One way of measuring such alignment is to challenge a model to take advantage of the learned embeddings and solve, possibly with additional supervision, tasks that require an understanding of real-world properties. Such tasks, often of practical interest, are defined \textit{downstream tasks}.

\vspace{0.1cm}\textbf{Pre-training for robust representation learning.}
In recent years, \textit{pre-training} proved a compelling way to learn meaningful embeddings. Especially when dealing with non-numerical inputs like images~\cite{krizhevsky2012imagenet} and text~\cite{howard-ruder-2018-universal}, pre-training demonstrated that it is possible to learn embeddings that automatically capture generic features and relationships of the raw data and can be exploited to solve many downstream tasks, with few (or even no) extra supervision~\cite{he2022masked, brown2020language, longpre2023flan}. Representation learning eliminates the need for \textit{feature engineering}, \ie to manually select or create relevant features. Instead, the model autonomously defines features directly from the raw input data.

\vspace{0.1cm}\textbf{Pre-training through pretext tasks.} 
Pre-training involves training the model on a series of \textit{self-supervised pretext tasks}. 
Unlike traditional end-to-end learning, this approach does not rely on externally provided labels.
Examples of pretext tasks include predicting the next word or phrase in a document~\cite{feder-etal-2022-causal} or reconstructing a masked patch of an image~\cite{he2022masked}. In both cases, the `correct label' is inherently derived from the input data itself. The component in charge of mapping the embedding space to the pre-text output is named \textit{Decoder}. The whole Encoder/Decoder architecture is trained on these self-supervised pretext tasks 
so to minimize the decoder error when compared to the correct data. 
Notice that, unlike traditional feature engineering -- where one explicitly models what they consider to be relevant aspects of the data -- pretext tasks encourage the model to independently uncover and understand which information to use to solve the task. The fundamental assumption here is that these features are indeed present, making it possible, for instance, to reconstruct a missing portion of an image given the remaining parts.
Ultimately, the literature agrees that the size and diversity of the pre-training dataset are key factors for its successful~\cite{devlin2018bert, kaplan2020scaling}. This is logical, as the embeddings shall capture generalizable aspects of the input data and avoid focusing on overly specific scenarios.

\vspace{0.1cm}\textbf{Leveraging the embeddings for downstream tasks.} A \textit{downstream task} refers to a specific problem a model is designed to solve, \eg the classification of samples into classes, as we consider here. 
Leveraging pre-trained architectures is an efficient approach for solving downstream tasks. The pre-trained encoder is first used to extract embeddings from the input data. A classification head is then added to use the knowledge captured by the encoder and perform the final classification.
The classification head, which ranges from a shallow model (\eg Random Forest (RF) or simple K-NN classifier) to a Neural Network (\eg Multi-Layer Perceptron (MLP)), works alongside the pre-trained encoder to form an overall \textit{classification model}.
Training this classification model for a specific downstream task requires labelled datasets for both train and test.
 
\vspace{0.1cm}\textbf{Frozen and unfrozen representation.} There are coarsely two options for training the classification model: (i) train only the classification head while keeping the pre-trained encoder \textit{frozen};
(ii) train the entire architecture end-to-end, with an \textit{unfrozen} pre-trained encoder.
The latter is often referred to as \textit{fine-tuning}, as it involves tailoring the general representations learned by the pre-trained model to address the particular task.
Although fine-tuning the entire model often yields better results, this is more computationally and memory-expensive~\cite {10.5555/3600270.3600279}. Additionally, when performed on a large amount of supervised data, end-to-end fine-tuning can significantly alter the encoder representation, potentially causing the model to forget its pre-trained knowledge and `overfit' to the downstream task. Sometimes, this can also lead the model to rely on tailored signals or \textit{shortcuts}.  Shortcuts are decision rules that perform well on standard benchmarks but do not transfer to more challenging testing conditions~\cite{geirhos2020shortcut} rather than robust features that truly represent the underlying task~\cite{McCoy2019RightFT}.

%% file: sections/03_Network_Repr_Learning/01_intro_repr_for_traff.tex
Researchers are exploring representation learning adoption to automatically learn meaningful representations of network traffic for tasks like traffic classification or QoE estimation. This section provides an overview of the most cited and recent approaches. The key characteristics of these solutions are summarized in Table~\ref{tab:summary_models}.

\begin{table*}
    \centering
    \resizebox{\textwidth}{!}{%
        \begin{tabular}{l|cccc|cccl}
        \toprule
        \multirow{2}{*}{\textbf{Model}} & \multicolumn{4}{c}{\textbf{Pre-training}} & \multicolumn{4}{|c}{\textbf{Downstream Classification}} \\ \cmidrule{2-9} 
         & \textbf{Architecture} & \textbf{Embedding Size} & \textbf{Task Types} & \textbf{Dataset} & \textbf{Cleaning} & \textbf{Split} & \textbf{\# Tasks } & \textbf{Datasets} \\ \midrule
        PacRep~\cite{meng2022packet} & BERT & 768 & None & Not needed & \color[HTML]{CB0000} Partial & \color[HTML]{CB0000} Packet & 6 & A, B, + \\
        PERT~\cite{he2020pert} & ALBERT & 768 & \color[HTML]{CB0000} MAE & $\neq$ & \color[HTML]{CB0000} No & Flow & \color[HTML]{CB0000}2 & \color[HTML]{CB0000} A, + \\
        ET-BERT~\cite{lin2022etbert} & BERT & 768 & \textcolor[HTML]{CB0000}{MAE}, SBP & \color[HTML]{CB0000} $\cap$ & \color[HTML]{CB0000} Partial & \color[HTML]{CB0000} Packet & 7 & A, B, C, + \\
        PTU~\cite{Peng2024PTU} & BERT & 768 & \textcolor[HTML]{CB0000}{MAE}, SSP, HIP, FIP & $\neq$ & \color[HTML]{CB0000} No & \color[HTML]{CB0000} Packet & 7 & A, B, C, + \\
        TrafficFormer~\cite{zhou2024trafficformer} & BERT & 768 & \textcolor[HTML]{CB0000}{MAE}, SODF & \color[HTML]{CB0000} $\cap$ & \color[HTML]{CB0000} Partial & \color[HTML]{CB0000} Packet & 6 & A, B, C, + \\
        netFound~\cite{guthula2023netfound} & BERT & 1024 & MAE & $\neq$ & \color[HTML]{CB0000} Partial & Flow & 5 & A, + \\
        YaTC~\cite{zhao_yet_2023} & ViT & 192 & \textcolor[HTML]{CB0000}{MAE} & \color[HTML]{CB0000} = & \color[HTML]{CB0000} No & \color[HTML]{CB0000} Unknown & 4 & A, B, + \\
        NetMamba~\cite{wang2024netmamba} & Mamba & 256 & \textcolor[HTML]{CB0000}{MAE} & \color[HTML]{CB0000} = & \color[HTML]{CB0000} Partial & Flow & 6 & A, B, + \\
        Pcap-Encoder & T5 & 768 & Autoencoder, Q\&A & $\neq$ & Full & Flow & 6 & A, B, C \\ \bottomrule
        \multicolumn{9}{r}{Datasets: A=ISCX-VPN, B=USTC-TFC, C=CSTNET-TLS1.3, +=other}
        \end{tabular}
    }
    \caption{Summary of representation learning models for traffic classification. Pitfalls are highlighted in red. }
    \label{tab:summary_models}
\end{table*}

%% file: sections/03_Network_Repr_Learning/02_generic_howto_repr_4_traff.tex
To adopt representation learning strategies, all proposed solutions follow a set of common steps which we analyse below, highlighting any potential pitfalls in the proposals\footnote{We based our discussion on the information in the papers and verified missing details using the authors' code and models when available.}.

\subsection{Choice of Model Architecture}
The first choice to make is whether to design a new encoder model or select an architecture previously presented in other areas. For traffic representation, all previous work builds on neural architectures presented in NLP or CV fields. 

\vspace{0.1cm}\textbf{Literature choices:} 
The motivation for using NLP-style models comes from the parallel between text, \ie sequences of characters organised in words, sentences, etc., and network traffic, \ie sequences of bytes organised in fields, packets, flows, etc.
\textit{PacRep}
~\cite{meng2022packet}, \textit{PERT}~\cite{he2020pert}, \textit{ET-BERT}~\cite{lin2022etbert}, \textit{TrafficFormer}~\cite{zhou2024trafficformer}, \textit{netFound}~\cite{guthula2023netfound} and \textit{PTU}~\cite{Peng2024PTU} use BERT-like models\cite{devlin2018bert, lan2019albert} borrowed from NLP and define network traffic specific pretext tasks for their training.

Other approaches draw inspiration from the field of image processing. They represent packets in the same flow as rows in a matrix to build an image. With this, they leverage tools such as \textit{Vision Transformer} (ViT)~\cite{dosovitskiy2021an} or \textit{Mamba}~\cite{gu2024mambalineartimesequencemodeling} to encode image-like inputs into the embedding space. \textit{YaTC}~\cite{zhao_yet_2023} and \textit{NetMamba}~\cite{wang2024netmamba} fall into this class.

\subsection{Pre-training Dataset}
For pre-training, collecting a large volume of unlabelled data is crucial to ensure comprehensive coverage of the main protocols. The commonly adopted strategy is to passively collect traces through network sniffers, and to leverage large publicly available datasets. 
Best practice suggests data used for pre-training to be much larger than those used for downstream task training -- as \textit{few-shot} learning should suffice if the learned representation is effective\footnote{For BERT, the ratio between supervised and self-supervised 
data samples ranges between 1:1,000 and 1:1,000,000.}. 
Additionally, the upstream and downstream datasets shall be different to limit model overfitting and data leakage.

\vspace{0.1cm}\textbf{Literature choices:} 
Different works can freely use different datasets for pre-training.
However, some studies employ the same datasets -- or part of them -- for both the pretraining and downstream tasks.
For example, \textit{ET-BERT} uses the \textit{ISCX-VPN}~\cite{DraperGil2016CharacterizationOE} and (likely) \textit{CSTNET-TLS1.3}~\cite{lin2022etbert} datasets for both tasks. Also, \textit{YaTC} and \textit{NetMamba} rely on the same datasets for upstream and downstream tasks training. 

\textbf{Associated pitfalls:} Dataset reuse is uncommon and discouraged in the CV and NLP domains. 
While for pre-training large datasets are strongly suggested, in the downstream task a large amount of data (likely) leads to forgetting -- especially when the representation model is unfrozen. 
Indeed, the encoder could override its pre-training knowledge and instead memorize task-specific patterns that can deceptively enhance the performance on downstream tasks. We will discuss this later in Sec.~\ref{sec:downstreamTraining}.

\subsection{Choice of Pre-training Tasks}
With pre-training, the model learns some generic and task-agnostic data patterns from the data themselves. Usual tasks require the encoder to develop predictive or reconstruction skills, often by masking part of the data or leveraging the data's temporal nature to predict future properties. 

\vspace{0.1cm}\textbf{Literature choices:} A common pre-training in networking involves \textit{reconstructing masked bytes} in a packet: The intuition behind this task is to encourage the model to identify correlations within the unmasked input to reconstruct missing parts. \textit{NetMamba}, 
\textit{YaTC}, \textit{netFound} 
and \textit{PERT} adopt this pre-training strategy, named \textit{Masked Autoencoder} (MAE)~\cite{he2022masked}.

\textit{ET-BERT} uses the original BERT pretext tasks of Masked Language Model (MLM) and Next Sentence Prediction (NSP). In \textit{ET-BERT} they call them \textit{Masked Burst Modelling} (MBM) -- a MAE-style task -- and \textit{Same-origin Burst Prediction} (SBP): given two packets, the model is queried whether the packets are part of the same burst\footnote{\textit{Burst}: sequence of consecutive packets that belong to the same flow.}.

\textit{TrafficFormer} keeps the first MAE task from \textit{ET-BERT}, but further complicates the second into \textit{Same Origin-Direction-Flow} (SODF): the model is not required to solve just a binary problem as in SBP, but also has to guess the direction, order, and corresponding flow of the packet.

\textit{PTU} also builds on \textit{ET-BERT} MAE, but adds a \textit{Same Session Prediction} (SSP) task, where the model has to predict whether two packets belong to the same session, and the \textit{Historical and Future Interval Prediction} (HIP and FIP) tasks to predict the time of arrival of previous and future packets in a flow.

\textit{netFound} uses the header information (e.g., packet length, \textit{TTL}, etc.) and the first 12 bytes of information in the payload, converts them into tokens, and then employs the standard MAE method for pre-training.

Finally, different from the others, \textit{PacRep} uses the off-the-shelf BERT model trained on text and does not design any network-specific pretext task.

\vspace{0.1cm}\textbf{Associated pitfalls:} Both linguistic and vision studies showed that words in a sentence and patches in an image exhibit significant correlations~\cite{saussure1916course, Chomsky_linguistic, vision_computational}. 
Contrarily, supposing a robust encryption algorithm, there is no correlation between the encrypted bytes in a packet payload. Hence, MAE tasks on encrypted payloads make little or no sense in the networking scenario. This is why more recent works, such as \textit{netFound}, only focus on unencrypted content during pre-training~\cite{guthula2023netfound}.

%% file: sections/pcapencoder.tex
\subsection{Mitigating the pitfalls: \tool}
\label{pcapencoder}

We propose a representation learning architecture that we explicitly design and train to automatically extract information from the protocol headers only -- that we assume still carry plain-text information. We call our proposed model architecture \textit{\tool}. \tool leverages two sequential pre-training phases that aim at capturing the contextual relationship among bytes and automatically extract the semantic of some packet header fields.

\begin{figure}[t]
    \centering
    \includegraphics[width=\columnwidth]{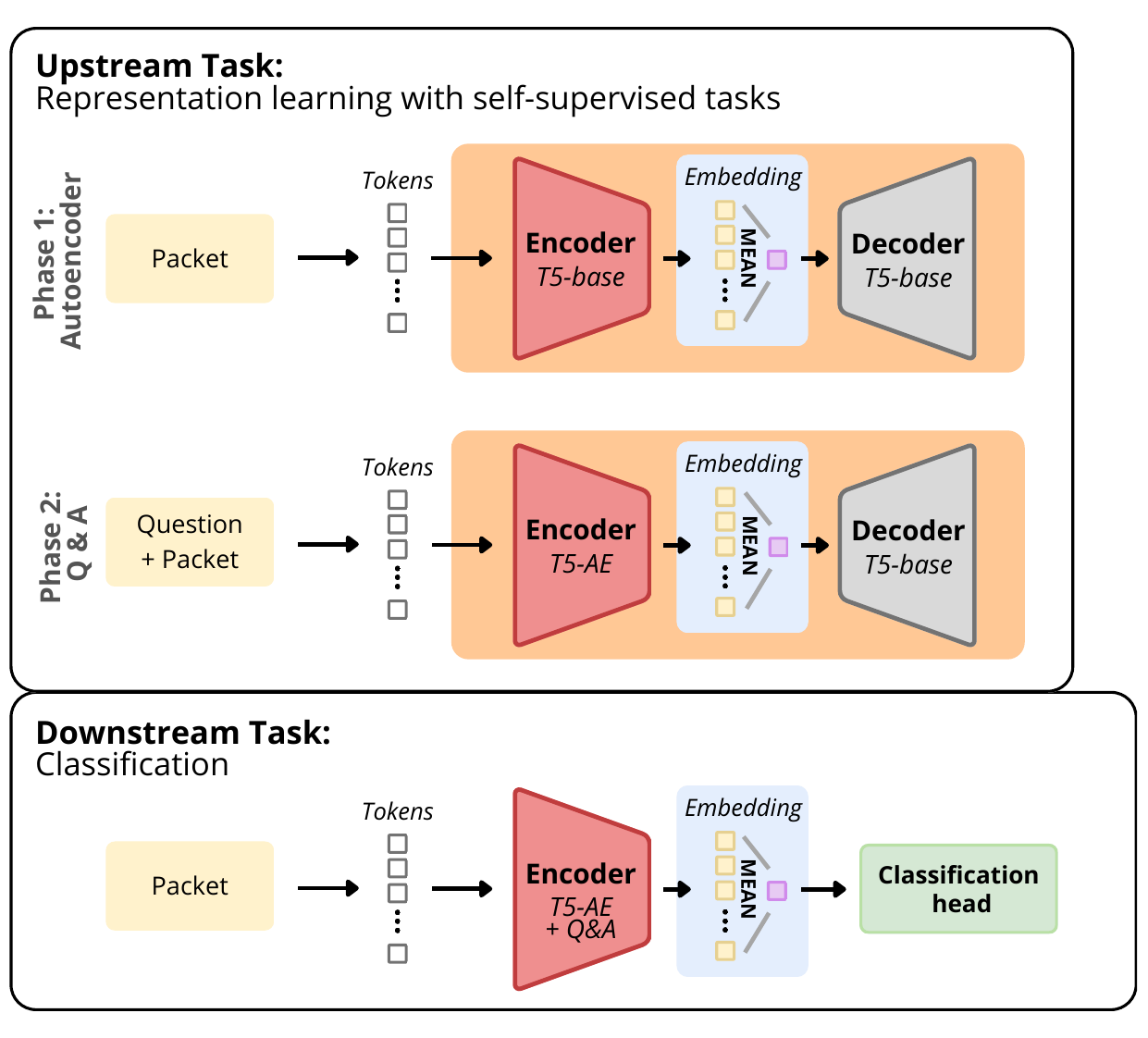}
    \caption{Schema of \tool, our proposal. }
	\label{fig:pipeline}
\end{figure}

The overall architecture is shown in Figure~\ref{fig:pipeline}. 
\tool is based on the well-known \textit{T5} (base) architecture~\cite{raffel2020exploring}, a sequence-to-sequence transformer-based model. T5 solves question-answering tasks: prompted a question and offered a context, the model learns how to answer.
For more details about the design of \tool, check Appendix \ref{sec:appendixEncoder} and our technical report \cite{DettoriTesi}.

\textbf{Phase 1: Encoder update.} We first update the original T5 encoder using raw packet traces to adapt it to the new data format and semantics. The encoder's goal is to map the original data into a numerical space, possibly removing redundant or useless (e.g., constant) information. For this, we train the (base) T5 to reconstruct the original packet from the internal representation.

T5 works using tokens to represent the input text. 
We convert each 2-byte long word into a hexadecimal number, separating each word by space. We feed this textual input to the original T5 tokenizer to generate tokens.
The encoder receives the packet divided into tokens and obtains a representation vector for each token. We modify the T5 encoder and add a bottleneck to obtain a single representation for the entire packet from the representations of its tokens. We test different architectures for this bottleneck and a simple \textit{mean pooling} layer suffices (see Appendix \ref{sec:appendixEncoder}). Finally, the decoder reconstructs the original packet tokens\footnote{To implement the encoder update, we use a dummy question and pass the packet over which the model operates.}. 

We start from the pre-trained \textit{T5-base} encoder and continue the training on traffic data with a standard cross-entropy loss function based on the difference between the predicted tokens and the actual ones. At the end, the encoder gives us a single representation of the input packet. We call this pre-trained model \textit{T5-AE}.
For this pre-training, we use MAWI~\cite{MAWI}, UNSW-NB15~\cite{UNSW-NB15} traces and a freshly collected trace from our campus\footnote{To ensure diversity and avoid the model memorising constant patterns, we randomize IP addresses and TTL values.}. This ensures both spatial and time diversity in the samples. Traces include IPv4 and IPv6 traffic, and mostly TCP, UDP and some ICMP traffic.
In total, we use $\approx1$GB of data or 500k packets.

\textbf{Phase 2: Question-answering.} Next, we fine-tune the \textit{T5-AE} model to extract the semantics of protocol headers. We task the model to answer questions related to some specific packet header or fields. 
In total, we define types of 8 questions, and create 50,000 instances of questions. 
We included (see Appendix \ref{sec:appendixEncoder}) retrieval questions applied across different protocols (TCP-IPv4/6, UDP-IPv4, and ICMP) and more complex questions that involve the computation on different fields.
Two examples of question prompts are `What is the destination IP address of the packet?' and `Is the packet's IP checksum correct?'.
We stress that we avoid pretext questions on the application payload (except its size), assuming encryption prevents any possible answer on the content.

This simple Q\&A training model allows one to pre-train \tool on specific protocol information. For instance, it can be tasked to find the SNI in the TLS handshake or learn where to find the A or AAAA record in DNS queries.

We start from the \textit{T5-AE} encoder trained at Phase 1 using the mean pooling bottleneck to represent each packet. For the question-answering task, the input consists of two parts: the query and the context. The query is the task we ask the model to solve. The context is the packet. Queries are in plaintext, so we use the original T5 tokenizer directly. For packets, we tokenize them as before.
At last, we separate the query from the context by the special token \texttt{</s>} that ensures the model correctly interprets the boundaries between the query and the context. For example: \textit{What is the time to live of the packet?}\texttt{</s>}\texttt{4500 4000 F7C6 ... CD19}.

We use the same traces as before for this second phase. Ablation studies on the \textit{T5-AE} and \textit{Q\&A} modules show that each component plays a beneficial role in enhancing model performance (details in Table \ref{tab:effectiveness_pretrain}).
At the end, we have a \textit{T5-AE+Q\&A} pre-trained model.  

\textbf{Downstream classifiers:}
As in other representation learning models proposed for traffic classification, we add a classification head made by a two-layer MLP with a ReLU activation function. It takes as input the \textit{T5-AE+Q\&A} embedding computed from the input packet. 
We train separated classifiers, one for each task. Depending on the number of classes in the task, we use binary cross-entropy or softmax as a loss function for the MLP. 

%% file: sections/04_Pipeline/pipeline.tex
In this section, we shift focus to the downstream task of the representation learning pipeline. As in the previous section, we explore the design choices of the most influential and recent approaches, highlighting potential pitfalls. Additionally, to provide a common ground for comparing results, we introduce a fair and effective benchmark pipeline to evaluate the representation learning capabilities of different models for network traffic classification tasks. Our benchmark standardizes the critical steps of trace gathering, cleaning, splitting, and sampling -- processes that are often inconsistently or incorrectly handled in the literature, making it difficult to fairly compare different methods.

\subsection{Dataset Preparation: Collecting, Cleaning, Splitting}\label{Dataset_and_Task_Definition}
We first discuss the choices on the datasets that previous authors used to train and evaluate their solutions on downstream tasks. The picture is extremely heterogeneous. Most authors use publicly available datasets, while a few autono\-mously collect traces~\cite{he2020pert, lin2022etbert, guthula2023netfound} -- some sharing (part of) them~\cite{lin2022etbert}. Each paper proposes a \textit{custom cleaning process} that removes spurious traffic (\eg ARP, DHCP, LAN-related protocols, etc.). Some remove flows or packets shorter than a given threshold~\cite{lin2022etbert, zhou2024trafficformer, wang2024netmamba, guthula2023netfound}. Some perform \textit{careless train-test splits}, ignoring the fact that packets of the same flow might leak information on the classification class~\cite{meng2022packet, lin2022etbert, Peng2024PTU}. In general, even when starting from the same dataset, all papers end up with a custom collection: even the number of classes per task often differs. 

We hereby call for a standardization in the following. The process we propose here can be extended to include other datasets and classification tasks.

\vspace{0.1cm}\textbf{Choice of dataset}: Instead of setting up specific data collection campaigns, we rely on previously used mainstream datasets and classification tasks created by the research community. These labelled datasets were generated through experiments conducted in controlled testbeds and offer a large data collection of encrypted traffic. We select three datasets among those commonly used in previous works, for a total of six tasks that we summarise in Table~\ref{tab:tasks}.  

$\bullet$ \textit{ISCX-VPN}~\cite{DraperGil2016CharacterizationOE}: This dataset contains traffic related to 6 different types of services (Web browsing, VoIP, Video Streaming, Chat, Email, P2P File transfer) using different applications (e.g., Chat with Skype or Hangouts), over plain or VPN-encrypted connection. We define three tasks: determining whether traffic is VPN-encrypted or not (\textit{VPN-binary}); Service classification (\textit{VPN-service}); and application classification (\textit{VPN-app}). 

$\bullet$ \textit{USTC-TFC}~\cite{wang_USTC-tfc}: This dataset contains a total of 20 applications, 10 are benign (BitTorrent, FaceTime, Gmail, Skype, ...) and 10 are malicious (malware run in controlled environments). We formulate two classification tasks: Malicious or not (\textit{USTC-binary}); and application classification (\textit{USTC-app}).

$\bullet$ \textit{CSTN-TLS1.3}~\cite{lin2022etbert}: This dataset contains a total of 120 classes, each referring to visits to a different TLS1.3-enabled website. The task here is to output the visited website (\textit{TLS-120}). The authors share only TCP flows from which they remove the TCP 3-way-handshake and the initial client TLS-Hello -- thus removing the plain-text SNI if present. This results in an `everything encrypted' payload scenario\footnote{In the original \textit{ET-BERT} paper the authors state the SNI is present, but in the public dataset it is not.}.

\begin{table}[]
\centering
\resizebox{\columnwidth}{!}{%
\begin{tabular}{c|c|c|c|c|c}
\toprule
\textbf{Dataset} & \textbf{Task} & \textbf{\#Class}  & \textbf{\#Train} & \textbf{\#Test} & \textbf{Description} \\ \midrule
ISCX-VPN & VPN-binary & 2 & 100,000 & 110,594 & Encrypted? \\
ISCX-VPN  & VPN-service & 6 & 120,000  & 111,368  & Voip, Chat, ... \\
ISCX-VPN & VPN-app  & 16 & 33,088 & 111,678  & Gmail, Vimeo, ... \\ \midrule
{\color[HTML]{000000} USTC-TFC}& USTC-binary  & 2  & 100,000  & 609,332  & Malware? \\
USTC-TFC  & USTC-app & 20& 69,680  & 609,477  & Gmail, Skype, ... \\ \midrule
CSTN-TLS1.3 & TLS-120 & 120 & 98,640 & 553,994 & 120 Websites \\ \bottomrule
\end{tabular}%
}
\caption{Downstream datasets and tasks.}
\label{tab:tasks}
\end{table}

\vspace{0.1cm}\textbf{Data cleaning:} Not supervising the trace collection process, data cleaning becomes a crucial step to ensure the quality and reliability of the datasets~\cite{flood2024bad}.
We summarize our interventions in the following four cases:

$\bullet$ \textit{Extraneous protocol filters}: Given the constraints of network data collection, certain extraneous protocols inevitably make their way into the datasets. Some traces include ARP, DHCP, broadcast protocols, etc. that question the definition of the classification task (\eg making predictions on ARP requests, which are not related to any classes). Some of the previous works~\cite{he2020pert, meng2022packet, zhao_yet_2023, Peng2024PTU} did not clean (or did not report how they cleaned) the traces, blindly trusting the data collection. Besides, \textit{netFound} retains traffic data related to the TCP, UDP and ICMP. For our benchmark, \textit{\uline{we define a superset of filters that we report in Appendix~\ref{filters} to filter out the irrelevant protocols to the classification task.}} ISCX and USTC traces contain 5\% and 10\% of spurious packets, respectively. CSTN is already filtered.

$\bullet$ \textit{Minimum size filters}: Some of the previous work filtered packets shorter than a minimum size~\cite{lin2022etbert, zhou2024trafficformer, wang2024netmamba, guthula2023netfound}. For example, in \textit{ET-BERT}, the authors remove all packets shorter than 80B\footnote{This filter is present in the code, but not mentioned in the paper.}. In \textit{TrafficFormer}, the authors remove flows shorter than 2kB or than three packets. In \textit{netFound}, the authors exclude flows with fewer than six packets and bursts containing two or fewer packets. Filtering based on packet or flow size alters the classification tasks since, for instance, all TCP signalling and acknowledgement packets could be ignored. Hence, \textit{\ul{we do not adopt and support filters based on minimum size/number.}}

$\bullet$ \textit{Classes support filters}: Some works limit the number of packets per class~\cite{lin2022etbert}, the number of flows per class~\cite{lin2022etbert, wang2024netmamba, zhou2024trafficformer}, or directly drop a class if the minimum support is not reached~\cite{lin2022etbert, wang2024netmamba, zhou2024trafficformer}. For example, \textit{TrafficFormer} discards classes with less than 10 flows and limits to 500 flows the others; \textit{ET-BERT} selects at most 5000 packets or 500 flows per class; \textit{NetMamba} discards rare classes and limits common ones, but authors do not report thresholds.
Since these filters alter the original data distribution, they change the nature of the problem compared to the initial dataset. The original datasets should reflect real-world conditions, while artificially modifying the underlying distributions introduces deviations that may impact the performance in practical deployments. \textit{\ul{We refrain from applying any class removal during testing. Differently, during training, researchers could use techniques that change the original class distribution, such as balancing the class samples (see next).}}

$\bullet$ \textit{Filters removing header information}: 
In an attempt to limit data leakage (\ie identifiers the model could memorize), some works propose to anonymise specific fields like IP addresses and TCP/UDP ports ~\cite{lin2022etbert, zhou2024trafficformer, Peng2024PTU, zhao_yet_2023, meng2022packet, wang2024netmamba}. For example, \textit{YaTC} randomises the IP address and sets the port number to zero; \textit{PacRep} and \textit{NetMamba} set both IP address and port to zero; \textit{PTU} removes IP address, MAC address and checksum; \textit{TrafficFormer} randomises IP address and ports, and some specific fields (such as timestamp) for data augmentation. \textit{ET-BERT} removes the IP header entirely. \textit{netFound} omits the explicit flow identifiers (\eg IP address, port, SNIs, etc), but keeps other header's information.

\textit{\ul{For the pre-training task, we consider it incorrect to remove any information from the header. In the downstream task, removing selected header fields compels the model to generalize across diverse network conditions}} -- for instance, it can no longer rely on memorizing that a specific IP range belongs to one server. \textit{\ul{It is hence part of the downstream model and training task design to hide some information (and force the model to generalise) or to leave such features and create a more dedicated model.}}

\vspace{0.1cm}\textbf{Dataset Splitting into Train and Test:} The golden rule in any ML pipeline is to avoid any leakage of information from the test set into the training/validation set. For traffic classification, we consider two basic splitting processes:

$\bullet$ \textit{Per-packet split:} Separate the packets based on their class; randomly split each class into train, validation and test sets.

$\bullet$ \textit{Per-flow split:} Separate the flows based on their class; randomly put all packets from the same flow into either train, validation and test sets.

\begin{figure}
    \centering
    \includegraphics[trim={0.8cm 1cm 0.8cm 0.5cm}, width=1\columnwidth]{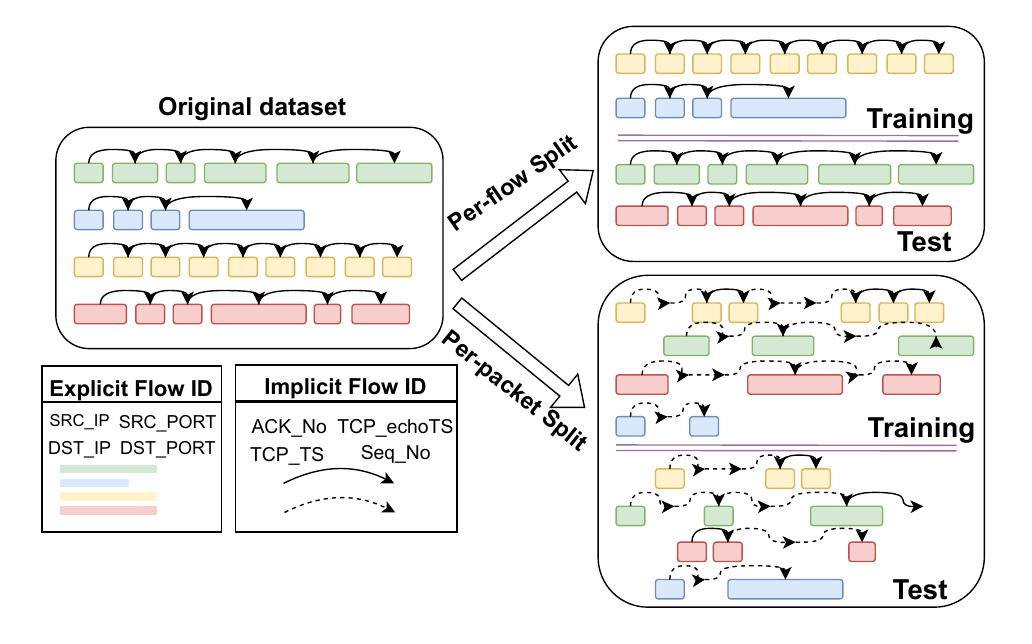}
    \caption{Per-flow and per-packet split.}
    \label{fig:split}
\end{figure}

Fig.~\ref{fig:split} sketches the differences between per-flow and per-packet split. Given flows, there are \textit{explicit flow identifiers} (ID) like the flow 5-tuple (represented by colours), and \textit{implicit flow IDs}, such as the TCP sequence (SeqNo) and ACK numbers (AckNo). For instance, TCP SeqNo and AckNo are randomly selected during the TCP three-way handshake, and all packets of the same flow then share values in a close range. Overall, the pairs (SeqNo, AckNo) create an implicit flow ID projecting all flow packets in a random space of $\approx$64 bits.
Similarly, the TCP timestamp option implicitly groups packets of the same session by a close-by timestamp.

All previous works adopt a per-packet split for packet classification tasks. Unfortunately, this leads to serious data leakage that allows the classifier to leverage implicit flow IDs to identify all packets of the same flow. Being the class of the flow available during training, the classifier can easily associate a packet to its flow, and then to its class.
Therefore, \textit{\ul{we propose the adoption of the per-flow split to remove simple implicit flow IDs that the classifier would not be able to leverage in real deployments.}}

Notice that more advanced splits are possible: per-session, per-client, per-location, per-time split, etc. Each stresses the ability of the model to generalise when transferred to other setups. Here, we limit our analysis to the basic per-packet and per-flow split.

\vspace{0.1cm}\textbf{Sampling a subset of the original dataset:} Different network applications generate different amounts of flows and packets. This translates into a class imbalance where a chatty application may exchange more packets than non-verbose applications, possibly creating an important class imbalance that challenges the downstream model training.
However, this imbalance is real: Limiting the number of packets or flows per class on the entire dataset, as in some of the previous works, introduces artifacts in the data distribution that could bias the model evaluation. 
\textit{\ul{Therefore, for the test set, we suggest not altering the class sample distribution as obtained from the previous step. }}
If for some reason (e.g., computational or time constraints at inference time) the test set must be reduced, \textit{\ul{we suggest \textit{stratified sampling}, which preserves the distribution of classes as in the collected data.}}

For the training (and validation) set, multiple valid choices can be made according to the used methodology. \textit{Balanced sampling} through oversampling or undersampling makes the number of samples of every class similar. This enables the model to better learn minority class samples and avoid the majority class dominating the model training process. Alternatively, one could use a weighted loss function or other ML techniques to compensate for class unbalance.

\subsection{Downstream classification}
\label{sec:downstreamTraining}

\textbf{Downstream task - Packet or flow:} Some work frames the final task as a packet classification problem~\cite{Peng2024PTU, meng2022packet}; others focus on flow classification~\cite{he2020pert, zhao_yet_2023, wang2024netmamba, guthula2023netfound, zhou2024trafficformer}; some does both~\cite{lin2022etbert}.
Here, we face both packet-level and flow-level tasks: given a packet (a flow), identify which class it belongs to. We define flow as the sequence of packets with the same 5-tuple, and consider both packets sent by the client and the server (bi-flow).
Finally, we consider all packets and flows to belong to the class the trace belongs to. For instance, when visiting a website, all cleaned packets and flows collected during such a visit inherit the same website label\footnote{Even if questionable, this is the same formulation previous works used.}.

\vspace{0.1cm}\textbf{Downstream model and training:} Each pre-trained encoder outputs an embedding, given some input sample. This tensor is the input to the classification head. In previous works, the classification head ranges from a simple MLP to a more complex transformer-based architecture (e.g. in \textit{PacRep}). Given the resulting classification model, authors trained it with simple supervision or using contrastive learning as in \textit{PacRep}. 
Given the complexity of the downstream task, \textit{\ul{one is free to choose the classification head model, but should keep in mind that the main learning effort lies in the self-supervised encoder -- so excessively complex heads are typically unnecessary and may not provide significant gains.}}

\vspace{0.1cm}\textbf{Pre-Trained Encoder - Frozen or unfrozen:} During the training of the final classifier, all previous works perform training in an end-to-end manner, i.e., the encoder architecture is \textit{unfrozen}.
While legitimate, this is in contrast with the idea that the representation produced by the encoder is actually representative. 
By freezing the encoding part of the model, the classification head should leverage the obtained generic representation.
Therefore, to check how meaningful the representation is, \textit{\ul{we advocate the usage of  \textit{frozen} encoder during the training of the downstream classifier.}}

\vspace{0.1cm}\textbf{Performance metrics: Accuracy and macro F1-Score.} 
\textit{Accuracy} measures the number of correct predictions. It is the ratio between the total number of correct predictions and the total number of predictions.
Accuracy treats all samples as equally important. In an unbalanced situation, it underweights the performance of minority classes.

\textit{Macro-averaged F1-Score} is the arithmetic mean (\ie unweighted mean) of all F1 scores per class. This metric equally weights errors across all classes, regardless of support.

All previous works but \textit{PacRep} and \textit{netFound} report the accuracy. Correctly, some\cite{lin2022etbert, Peng2024PTU, zhou2024trafficformer, he2020pert} present the macro F1-Score too. \textit{YaTC}, \textit{NetMamba} and \textit{netFound}\footnote{Detail inferred from the code.} misleadingly use the micro F1-Score -- which favours majority classes; \textit{PacRep} only reports micro F1 and macro F1 scores.

\textit{\ul{To evaluate the performance of the classifier, we suggest using both the accuracy and the macro-averaged F1 score.}}

%% file: sections/05_Results/results.tex
Here we describe the experimental setup to compare and understand the real potential of representation learning for the traffic classification tasks.

\begin{table*}[!t]
\centering
\resizebox{\textwidth}{!}{%
    \begin{tabular}{l|cc|cc|cc|cc|cc|
    >{\columncolor[HTML]{FFFFFF}}c 
    >{\columncolor[HTML]{FFFFFF}}c}
    \toprule
    \multicolumn{1}{c|}{} & \multicolumn{2}{c|}{\textbf{VPN-binary (2)}} & \multicolumn{2}{c|}{\textbf{VPN-service (6)}} & \multicolumn{2}{c|}{\textbf{VPN-app (16)}} & \multicolumn{2}{c|}{\textbf{USTC-binary (2)}} & \multicolumn{2}{c|}{\textbf{USTC-app (20)}} & \multicolumn{2}{c}{\textbf{TLS-120}} \\ \cmidrule{2-13} 
    \multicolumn{1}{c|}{\multirow{-2}{*}{\textbf{\begin{tabular}[c]{@{}c@{}}Model\\ (Per-flow split)\end{tabular}}}} & \multicolumn{1}{c}{AC} &  F1 & \multicolumn{1}{c}{AC} &  F1 & \multicolumn{1}{c}{AC} &  F1 & \multicolumn{1}{c}{AC} &  F1 & \multicolumn{1}{c}{AC} &  F1 & \multicolumn{1}{c}{AC} &  F1 \\ \midrule
    \textbf{ET-BERT} & 84.7 & 84.6 & 71.7 & 64.2 & {\color[HTML]{000000} 59.2} & {\color[HTML]{CB0000} 43.7} & 100.0 & 100.0 & 84.9 & 79.6 & {\color[HTML]{CB0000} 10.9} & {\color[HTML]{CB0000} 6.7} \\ 
    \textbf{YaTC} & 83.9 & 83.9 & 69.2 & 60.1 & {\color[HTML]{000000} 60.9} & {\color[HTML]{CB0000} 44.3} & 99.5 & 99.5 & 85.2 & 78.0 & {\color[HTML]{CB0000} 15.5} & {\color[HTML]{CB0000} 9.6} \\ 
    \textbf{NetMamba} & 75.0 & 74.5 & 56.9 & {\color[HTML]{CB0000} 49.0} & {\color[HTML]{CB0000} 39.6} & {\color[HTML]{CB0000} 28.4} & 97.6 & 97.5 & 72.5 & 57.7 & {\color[HTML]{CB0000} 8.8} & {\color[HTML]{CB0000} 4.5} \\ 
    \textbf{TrafficFormer} & 90.9 & 90.9 & 76.5 & 69.4 & 67.7 & 54.4 & 100.0 & 100.0 & 72.0 & 65.0 & {\color[HTML]{CB0000} 29.7} & {\color[HTML]{CB0000} 24.0} \\
    \textbf{netFound} & 76.0 & 61.9 & {\color[HTML]{CB0000} 47.3} & {\color[HTML]{CB0000} 36.5} & {\color[HTML]{CB0000} 32.9} & {\color[HTML]{CB0000} 15.3} & 99.4 & 99.4 & 58.0 & {\color[HTML]{CB0000} 30.7} & {\color[HTML]{CB0000} 1.9} & {\color[HTML]{CB0000} 0.5}\\ \midrule
    \textbf{\tool} & \textbf{99.9} & \textbf{99.9} & \textbf{92.1} & \textbf{89.8} & \textbf{83.5} & \textbf{71.0} & \textbf{100.0} & \textbf{100.0} & \textbf{91.0} & \textbf{87.1} & \textbf{71.0} & \textbf{63.7} \\ \bottomrule
    \end{tabular}%
}
\caption{Results of \tool and the three SoA models for packet classification. Per-flow split, Frozen encoders. We report accuracy (AC) and macro F1-Score (F1). Results below 50\% are highlighted in red, best in bold.} 
\label{tab:packet_class_cleaned}
\end{table*}

\vspace{0.1cm}\textbf{Downstream models:} For our experiments, we select five representative models: \textit{ET-BERT},
\textit{YaTC},
\textit{NetMamba}, \textit{TrafficFormer} and \textit{netFound}.
They use different representation learning models: BERT, ViT and Mamba.
We download the pre-trained models from each original repository.
We add \tool,
and compare against shallow models (without representation learning) as baselines.

For each model, we follow the data preparation and hyperparameters suggested by the original papers when available. Refer to Appendix~\ref{sec:appendixHyperparameters} for a detailed overview. When facing \textit{packet classification} with the flow-embedders, we \textit{Repeat} the same packet 5 times to form an artificial flow and get the resulting embeddings in output\footnote{We also test a \textit{Padding} strategy where 4 padding packets with all zeros follow the packet. The Repeat strategy offers better results.}. For \textit{netFound}, we fill its maximum input (72 packets) with the same packet and pad the multimodal information (packet direction, packet interval, and etc.) with the same value or zero. We use the original flow-embedder models for the \textit{flow classification task}. For \tool, we consider a simple majority voting on the classification of the first 5 packets of each flow.

\vspace{0.1cm}\textbf{Downstream task evaluation:} As reported in Section~\ref{sec:training_pipeline}, we split each dataset into training and test partitions, according to a 7:1 ratio. We adopt the \textit{per-flow split} strategy. We make sure that long flows are evenly distributed within the partitions. To stress the few-shot learning abilities of the models, in the training set, we \textit{balance} classes by undersampling each class to the minority class. For flows longer than 1,000 packets, we randomly select 1,000 packets across the flow.
Table~\ref{tab:tasks} reports the number of samples in the training and testing sets. The proportion between the train and test samples changes for each downstream dataset due to the undersampling of the training part.

We perform a K-Fold cross-validation of the training partition with $K=3$ (2/3 used for training, 1/3 for validation, 3 folds). 
In all experiments, we test all models and configurations with the exact same splits for a fair comparison. 

To compare with the \textit{per-packet split} scenario, we create a second split simply following an 8:1:1 random split into train, validation and test -- as originally proposed in \textit{ET-BERT}. In this case, packets from the same flow can end up in both trains and tests.

\section{Results}\label{sec:results}
We perform all the experiments HPC Cluster equipped with NVIDIA Tesla V100 SXM2 GPUs. We use Python and Pytorch for the implementation and make all datasets, models, and code available to the community for reproducibility and to foster further studies.

\subsection{Packet-Level Traffic Classification}
We start from the proposed flow-split-based scenario with frozen encoders on the packet classification tasks. 

\vspace{0.1cm} \textbf{Per-Flow Split -- Frozen encoder:} We consider the six tasks we presented in Section~\ref{sec:training_pipeline} and test all representation learning models with the per-flow split and frozen encoders. We report results in Table~\ref{tab:packet_class_cleaned}. Surprisingly, the performance of all models is very poor, up to 80\% lower than that reported in their respective papers. Only in the simplest binary classification tasks, \textit{VPN-binary} and \textit{USTC-binary}, all models offer solid results. 
On the most challenging \textit{VPN-app} and \textit{TLS-120} tasks, the performance is really disappointing.

Notice how \tool performs best in all tasks, significantly outperforming other methods. Yet, in the most complex tasks, it struggles to achieve excellent results. 

Although it contrasts with previous studies, the poor performance in encrypted scenarios is justified by the debatable assumption made by previous work according to which some information can still be extracted from the encrypted payload. Since these models are designed to disregard information from packet headers, they rely on minimal data, primarily packet direction and size.
In contrast, \tool provides an informative representation to the classification head, enabling it to distinguish the application that generated a packet from the network and the transport headers summarised by \tool.

Given that some tasks are very simple, in the following, we dismiss three and focus only on the \textit{VPN-app} and \textit{TLS-120} tasks.

\vspace{0.1cm} \textbf{Per-Flow Split -- Unfrozen Encoder:} To investigate the root cause of poor performance, we let the classification model fine-tune the embedder part as well, \ie we unfreeze the encoder.
We report results in Table~\ref{tab:flow_split_frozen_vs_unfrozen} for \textit{VPN-app} and \textit{TLS-120} cases. As expected, all proposed models improve their performance. Still, unfreezing the encoder does not suffice to reach a satisfactory result for none of them. 
\tool still achieves the best performance: Notice that the unfreezing boost is less significant than the counterparts -- about 5\% improvement only. This confirms that \tool still relies on its pre-training knowledge and does not require re-learning with end-to-end training. 

\begin{table}[!t]
\centering
\resizebox{\columnwidth}{!}{%
    \begin{tabular}{c|cc|cc|cc|cc}
    \toprule
     & \multicolumn{4}{c|}{\textbf{VPN-app (16)}} & \multicolumn{4}{c}{\textbf{TLS-120}} \\ \cmidrule{2-9} 
     & \multicolumn{2}{c}{\textbf{Frozen}} & \multicolumn{2}{c|}{\textbf{Unfrozen}} & \multicolumn{2}{c}{\textbf{Frozen}} & \multicolumn{2}{c}{\textbf{Unfrozen}} \\ \cmidrule{2-9} 
    \multirow{-3}{*}{\textbf{\begin{tabular}[c]{@{}c@{}}Model\\ (Per-flow split)\\\,\end{tabular}}} & \multicolumn{1}{|c}{AC} & F1 & \multicolumn{1}{|c}{AC} & F1 & \multicolumn{1}{|c}{AC} & F1 & \multicolumn{1}{|c}{AC} & F1 \\ \midrule
    \textbf{ET-BERT} & 59.2 & {\color[HTML]{CB0000} 43.7} & 82.8 & 69.7 & {\color[HTML]{CB0000} 10.9} & {\color[HTML]{CB0000} 6.7} & {\color[HTML]{CB0000} 28.0} & {\color[HTML]{CB0000} 21.5} \\
    \textbf{YaTC} & 60.9 & {\color[HTML]{CB0000} 44.3} & 79.1 & 65.2 & {\color[HTML]{CB0000} 15.5} & {\color[HTML]{CB0000} 9.6} & {\color[HTML]{CB0000} 36.6} & {\color[HTML]{CB0000} 31.4} \\ 
    \textbf{NetMamba} & {\color[HTML]{CB0000} 39.6} & {\color[HTML]{CB0000} 28.4} & 80.4 & 65.9 & {\color[HTML]{CB0000} 8.8} & {\color[HTML]{CB0000} 4.5} & {\color[HTML]{CB0000} 40.7} & {\color[HTML]{CB0000} 35.3} \\ 
    \textbf{TrafficFormer} & 67.7 & 54.4 & 73.1 & 61.0 & {\color[HTML]{CB0000} 29.7} &  {\color[HTML]{CB0000} 24.0} & {\color[HTML]{CB0000} 43.7} & {\color[HTML]{CB0000} 38.9} \\
    \textbf{netFound} & {\color[HTML]{CB0000} 32.9} & {\color[HTML]{CB0000} 15.3} & 70.4 & 57.3 & {\color[HTML]{CB0000} 1.9} & {\color[HTML]{CB0000} 0.5} & {\color[HTML]{CB0000} 39.9} & {\color[HTML]{CB0000} 35.2} \\\midrule
    \textbf{\tool} & \textbf{83.5} & \textbf{71.0} & \textbf{85.6} & \textbf{74.8} & \textbf{71.0} & \textbf{63.7} & \textbf{77.3} & \textbf{69.2} \\ \bottomrule
    \end{tabular}%
}
\caption{Per-flow split, frozen and unfrozen encoder. Results improve, but models still struggle in challenging setups.}
\label{tab:flow_split_frozen_vs_unfrozen}
\end{table}

\vspace{0.1cm} \textbf{Per-Packet Split -- Frozen vs Unfrozen Encoder:} 
Wondering why the performance does not yet saturate to the promised good performance, we observe what happens with the per-packet split as originally adopted in all previous works. We present results in Table~\ref{tab:packet_split_frozen_vs_unfrozen} for both frozen and unfrozen encoder setups on the two most challenging tasks.

\begin{table}[]
\centering
\resizebox{\columnwidth}{!}{%
    \begin{tabular}{c|cccc|cccc}
    \toprule
     & \multicolumn{4}{c|}{\textbf{VPN-app (16)}} & \multicolumn{4}{c}{\textbf{TLS-120}} \\ \cmidrule{2-9} 
     & \multicolumn{2}{c}{\textbf{Frozen}} & \multicolumn{2}{c|}{\textbf{Unfrozen}} & \multicolumn{2}{c}{\textbf{Frozen}} & \multicolumn{2}{c}{\textbf{Unfrozen}} \\ \cmidrule{2-9} 
    \multirow{-3}{*}{\textbf{\begin{tabular}[c]{@{}c@{}}Model\\ (Per-packet split)\\\,\end{tabular}}} & \multicolumn{1}{|c}{AC} & F1 & \multicolumn{1}{|c}{AC} & F1 & \multicolumn{1}{|c}{AC} & F1 & \multicolumn{1}{|c}{AC} & F1 \\ \midrule
    \textbf{ET-BERT} & 69.5 & \multicolumn{1}{c|}{64.7} & 96.8 & 97.0 & {\color[HTML]{CB0000} 17.5} & \multicolumn{1}{c|}{{\color[HTML]{CB0000} 10.2}} & 97.4 & 96.8 \\ 
    \textbf{YaTC} & 73.2 & \multicolumn{1}{c|}{67.7} & \textbf{98.5}  & \textbf{98.5} & {\color[HTML]{CB0000} 26.8} & \multicolumn{1}{c|}{{\color[HTML]{CB0000} 17.7}} & \textbf{98.2}  & \textbf{97.7}  \\ 
    \textbf{NetMamba} & 53.5 & \multicolumn{1}{c|}{{\color[HTML]{CB0000} 45.1}} & 98.4  & 98.4  & {\color[HTML]{CB0000} 13.5} & \multicolumn{1}{c|}{{\color[HTML]{CB0000} 5.3}} & 97.4 & 96.8 \\ 
    \textbf{TrafficFormer} & 87.5 & \multicolumn{1}{c|}{85.6} & 95.6  & 95.2  & 53.3 & \multicolumn{1}{c|}{{\color[HTML]{CB0000} 48.2}} & 86.0 & 83.3 \\ 
    \textbf{netFound} & {\color[HTML]{CB0000} 35.3} & \multicolumn{1}{c|}{{\color[HTML]{CB0000} 18.7}} & 89.6  & 89.0  & {\color[HTML]{CB0000} 10.8} & \multicolumn{1}{c|}{{\color[HTML]{CB0000} 2.3}} & 71.0 & 67.4 \\ \midrule
    \textbf{\tool} & \textbf{91.9} & \multicolumn{1}{c|}{\textbf{90.6}} & 94.3  & 93.9 & \textbf{85.7} & \multicolumn{1}{c|}{\textbf{81.0}} & 88.6 & 80.3 \\ \bottomrule
    \end{tabular}
}
\caption{Per-packet split scenario. Eventually, in this wrong settings and unfrozen encoder, performance reaches the promised $>90\%$ accuracy.}
\label{tab:packet_split_frozen_vs_unfrozen}
\end{table}

First, focus on the frozen encoder scenario. The performance that all previous models achieve is still far from satisfactory. In particular, in the \textit{TLS-120}, we obtain very poor results.
Compare now with an unfrozen encoder: The accuracy finally largely improves, reaching up to $98\%$.

Two main take-home messages arise:
\textit{\ul{First, per-packet split enables some data leakage that allows the classifier to finally find exploitable patterns.
Second, the representation offered by the frozen encoder is not useful for solving classification tasks.}} Only when end-to-end training is enabled, the models learn patterns that allow them to shine.

\begin{figure}[t]
    \centering
    \includegraphics[trim={0.2cm 0.2cm 0.2cm 0.2cm},clip, width=\columnwidth]{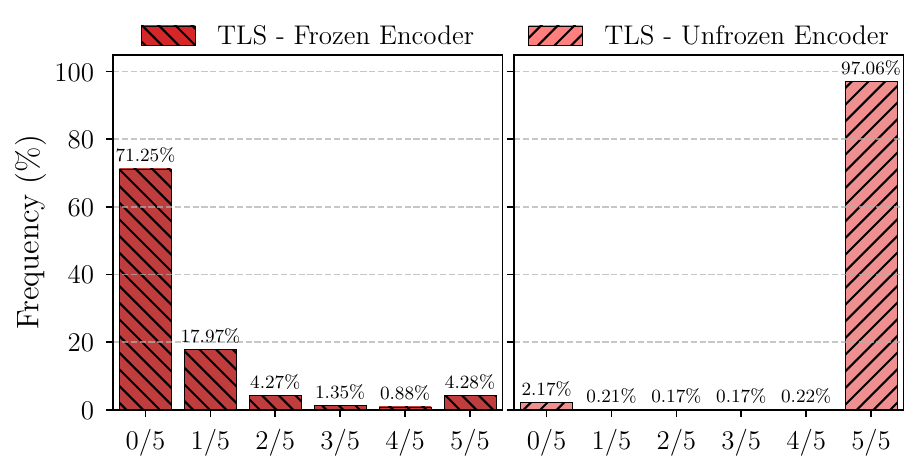}
    \caption{5-NN purity of embeddings for \textit{ET-BERT}. With a frozen encoder, $71\%$ of points do not have a sample of the same class as TOP-5 neighbour. Situation changes only when the encoder is unfrozen.}
    \label{fig:knn_plot}
\end{figure}

To gauge the representativeness of the embeddings, we compute, for each packet, the number of neighbouring packets that have the same class, using their representation in the embedding space. The intuition is that if the embedder can project packets with the same class in the same portion of the embedding space, most of the packet's neighbours should be of the same class. For each point, we consider the 5-nearest neighbours (5-NN) and count how many samples are of the correct class (5-NN purity). Fig.~\ref{fig:knn_plot} shows the analysis considering \textit{ET-BERT} embeddings in frozen (left) and unfrozen (right) setups, in the \textit{TLS-120} case. Given a packet of class $c$, in the former case, 71\% of packets have no neighbour of class $c$. After fine-tuning, the embedding changes drastically, so that now 97\% of packets have all the 5 neighbours of class $c$. The original embeddings lack meaningful information, and it is only during end-to-end fine-tuning that the classification model adapts them to specifically address the downstream tasks. The same holds for the embeddings of other models, not reported here for the sake of brevity. In a nutshell, the models need to modify all their weights to solve the classification task, as their original representations are uninformative.

Unfortunately, the patterns activated by the per-packet split are misleading and impractical for real-world use. The per-packet split strategy suffers from severe data leakage, allowing packets from the same flow to appear in both the training and test sets. As a result, the model learns information it should not rely on, i.e., it relies on shortcuts. Specifically, some implicit flow IDs enable the model to link a test packet to its corresponding flow and, ultimately, its class.

\begin{table}[]
\centering
\resizebox{.8\columnwidth}{!}{%
\begin{tabular}{c|c|cc}
\toprule
\textbf{Scenario} & \textbf{Dataset} & \textbf{AC} & \textbf{F1}  \\ \midrule
 & Original & \textbf{97.4} & \textbf{96.8} \\ \cmidrule{2-4} 
 & \begin{tabular}[c]{@{}c@{}}w/o SeqNo/AckNo\\ w/o Timestamp\\ (only test)\end{tabular} & {\color[HTML]{CB0000} 19.5} & {\color[HTML]{CB0000} 15.4} \\ \cmidrule{2-4} 
\multirow{-6}{*}{
\begin{tabular}[c]{@{}c@{}}\textbf{Per-packet Split}\end{tabular}
} & \begin{tabular}[c]{@{}c@{}}w/o SeqNo/AckNo\\ w/o Timestamp\\ (train + test)\end{tabular} & 52.2 & {\color[HTML]{CB0000} 48.2} \\ \cmidrule{2-4} 
  & w/o Pre-training & 97.1 & 96.4 \\ \midrule
\begin{tabular}[c]{@{}c@{}}\textbf{Per-flow Split}
\end{tabular} & Original & {\color[HTML]{CB0000} 28.0} & {\color[HTML]{CB0000} 21.5} \\ \bottomrule
\end{tabular}%
}
\caption{Impact of implicit flow ID on the unfrozen \textit{ET-BERT} and ablation study on the pre-training strategy. }
\label{tab:ablation_study_bert}
\end{table}

\vspace{0.1cm} \textbf{Removing implicit flow IDs:}
We systematically analyse the cause of the performance increase with the per-packet split and unfrozen model. We focus on \textit{ET-BERT} as a case study, with the \textit{TLS-120} task. Table~\ref{tab:ablation_study_bert} summarises these experiments.
The first row reports (as in Table \ref{tab:packet_split_frozen_vs_unfrozen}) the excellent performance with the per-packet split and unfrozen encoder. 

Now, consider the same model tested on the same data but with randomized SeqNo, AckNo, and TCP timestamps (second row). The results drop by approximately 80\%. This suggests that the model relies on these shortcuts during training, leading to an abrupt performance decline when those shortcuts are unavailable.

Train the same model now on a training set where the implicit flow IDs are removed -- and no easy shortcuts are present. Test this model on a test set without shortcuts too. Results improve. That is, the model -- trained unfrozen -- looks for other patterns that still allow it to solve the classification task, even if in an unsatisfactory manner. Some data leakage is still present.

To check the effectiveness \textit{ET-BERT} pre-training strategy, we run an experiment in which we destroy the pre-training knowledge by initialising the ET-BERT model weights with random values. We then fine-tune this untrained model to solve the TLS-120 downstream task. We report results in Table~\ref{tab:ablation_study_bert} in the w/o Pre-training row. Results are on par with those of the pre-trained \textit{ET-BERT} model. This strongly suggests that the \textit{ET-BERT} pre-training is mostly useless.

Lastly, in the last row we report the results in a per-flow split (the same as Table \ref{tab:flow_split_frozen_vs_unfrozen}). Here, packets are naturally and consistently separated into train or test sets, and the model has a hard time finding easy patterns.

In short, the per-packet split introduces dangerous information leaks, allowing the model to find easy patterns that will not be available during deployment. It is crucial to carefully split the data to avoid this issue: A per-flow split helps mitigate the problem. 

\vspace{0.1cm} \textbf{Ablation study on \tool:} we perform an ablation study on \tool in the per-flow split scenario when (i) removing the IPs, (ii) removing the TCP/IP header entirely, or (iii) removing the application payload. Changes are applied in both the train and test sets, and the encoder is kept frozen during training.

We report these results in Table~\ref{tab:ablation_pcapEncoder}. The performance reduces by removing the IP Addresses, and collapses by removing the entire IP and TCP headers. This is expected given that \tool is designed to ignore the payload. In fact, removing the application layer payload has a limited (\textit{VPN-app}) or no impact (\textit{TLS-120} -- everything encrypted scenario). By design -- and in practice -- the encrypted payload cannot make any significant contribution to the classification task.

\begin{table}[t]
\resizebox{0.7\columnwidth}{!}{%
\begin{tabular}{ccc}
\hline
\textbf{\begin{tabular}[c]{@{}c@{}}Model\\ (Per-flow split)\end{tabular}} & \multicolumn{1}{l}{\textbf{VPN-app (16)}} & \multicolumn{1}{l}{\textbf{TLS-120}} \\ \hline
\textbf{w/o IP addr.} & 52.5 & {\color[HTML]{CB0000}13.0} \\
\textbf{w/o header}   & {\color[HTML]{CB0000}16.4} & {\color[HTML]{CB0000}1.5}  \\
\textbf{w/o payload}  & 66.7 & 63.6 \\ \hline
\textbf{base}         & 71.0 & 63.7 \\ \hline
\end{tabular}%
}
\caption{Ablation Study on \tool in the flow-based split scenario when removing the IPs, headers and payloads (Macro F1-Scores).}
\label{tab:ablation_pcapEncoder}
\end{table}

\begin{table}[]
\centering
\resizebox{\columnwidth}{!}{%
\begin{tabular}{c|ccc|ccc}
\hline
 & \multicolumn{3}{c|}{\textbf{VPN-app (16)}} & \multicolumn{3}{c}{\textbf{TLS-120}} \\ \cline{2-7} 
\multirow{-2}{*}{\textbf{\begin{tabular}[c]{@{}c@{}}Model\\ (Per-flow split)\end{tabular}}} & \multicolumn{2}{c}{\textbf{base}} & \multicolumn{1}{l|}{\textbf{w/o IP addr}} & \multicolumn{2}{c}{\textbf{base}} & \multicolumn{1}{l}{\textbf{w/o IP addr}} \\ \hline
\textbf{RF} & \multicolumn{2}{c}{{\color[HTML]{000000} 81.1}} & 72.4 & \multicolumn{2}{c}{78.0} & {\color[HTML]{FE0000} 39.4} \\
\textbf{XGBoost} & \multicolumn{2}{c}{82.1} & 73.2 & \multicolumn{2}{c}{82.0} & {\color[HTML]{FE0000} 41.3} \\
\textbf{LightGBM} & \multicolumn{2}{c}{82.6} & 74.5 & \multicolumn{2}{c}{82.4} & {\color[HTML]{FE0000} 40.6} \\
\textbf{MLP} & \multicolumn{2}{c}{65.1} & 52.5 & \multicolumn{2}{c}{68.8} & {\color[HTML]{FE0000} 30.5} \\ \hline
\end{tabular}%
}
\caption{Macro F1-Scores of ML baselines. We try the baseline with and without the IP information.}
\label{tab:ml_vs_tool}
\end{table}

\vspace{0.1cm} \textbf{\tool vs Shallow models:}
So far, \tool has proved to offer the best performance in the realistic per-flow split scenario. But how does it compare with traditional ML approaches?

In Table~\ref{tab:ml_vs_tool}, we compare the performance of some shallow models with that of \tool. We report the F1 scores when provided the same input as \tool (\textit{base} settings). We manually select which packet header field to use as features for training and testing (details are in Appendix~\ref{shallow_models_detail}). The results show that the shallow ML models perform better than \tool. This happens because \tool relies on pre-training tasks to autonomously extract useful features from the raw byte stream, whereas we provide shallow models with custom features extracted from significant protocol fields pre-selected by networking experts.

Focus on the \textit{w/o IP addr} column in Table~\ref{tab:ml_vs_tool} that shows what happens when we remove the IP address from the data. The performance of shallow models drops, particularly in the TLS-120 task. With no access to IP addresses, the Random Forest relies on other fields, which overall provide less information. However, even in this configuration, the use of handcrafted features still enables the shallow models to outperform \tool.

\begin{figure}[t]
    \centering
    \includegraphics[width=1\linewidth]{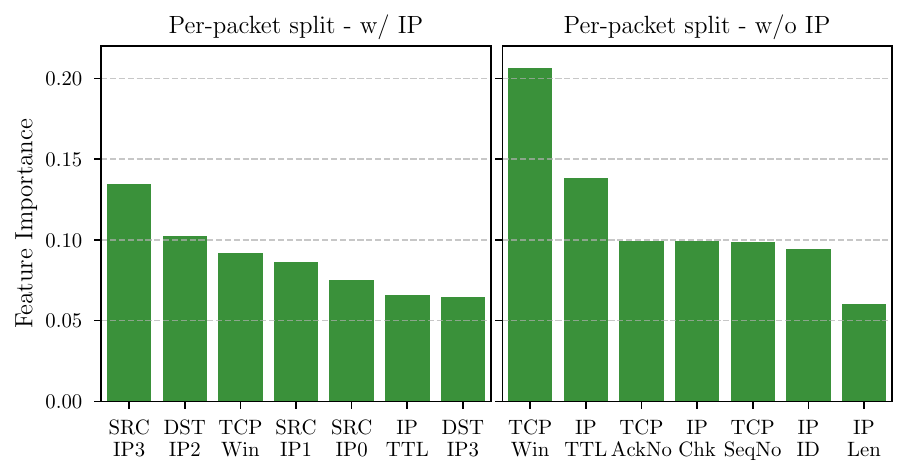}
    \caption{Feature importance for the Random Forest model on the per-packet split for the TLS-120 problem. We both examine the case with and without IP among input features.}
    \label{fig:FeatImportance}
\end{figure}

\vspace{0.1cm} \textbf{Shallow Models -- Feature Importance:} We leverage the feature importance scores provided by the Random Forest to identify the features the model relies on for its decisions. Fig.~\ref{fig:FeatImportance} reports the scores for the TLS-120 problem. We consider the per-packet split scenario to highlight the shortcuts on explicit and implicit flow IDs. 

Observe the left plot: this is the base scenario. All headers' features are available for training. IP addresses are explicit flow IDs and the model reaches an accuracy of $98.9$\%. The most relevant features are the different octets of the source and destination IP addresses (see \textit{SRC IP3}, \textit{DST IP2}, \textit{SRC IP1}, \textit{SRC IP0} and \textit{DST IP3})\footnote{Recall that we use bi-flows, and the source/destination IP addresses can correspond to both the client and server.}.

In the right plot, we report results after removing IP addresses from the input features. Here, sequence and acknowledgment numbers -- implicit flow identifiers -- become the most important features. Despite the removal of explicit identifiers, the model still achieves high performance with an accuracy of $92.6$\%, reflecting the flaws of the per-packet split setup.

\subsection{Flow-Level Traffic Classification Results} 
At last, given that \textit{YaTC}, \textit{NetMamba}, \textit{TrafficFormer} and \textit{netFound} models were designed for flow representation, we compare how they perform in a flow-level network traffic classification task. We take the same two challenging datasets used in the previous section and keep all flows that have at least 5 packets (the best case for encoders). Only per-flow split is viable in this case. 
We train the classification model using both frozen and unfrozen encoders using the same 3-fold approach. 

\begin{table}[]
\centering
\resizebox{.9\columnwidth}{!}{%
    \begin{tabular}{c|cccc|cccc}
    \toprule
     & \multicolumn{4}{c|}{\textbf{VPN-app (16)}} & \multicolumn{4}{c}{\textbf{TLS-120}} \\ \cmidrule{2-9} 
     & \multicolumn{2}{c}{\textbf{Frozen}} & \multicolumn{2}{c|}{\textbf{Unfrozen}} & \multicolumn{2}{c}{\textbf{Frozen}} & \multicolumn{2}{c}{\textbf{Unfrozen}} \\ \cmidrule{2-9} 
    \multirow{-3}{*}{\textbf{\begin{tabular}[c]{@{}c@{}}Model\\ (Per-flow split)\\\,\end{tabular}}} & AC & \multicolumn{1}{c|}{F1} & AC & F1 & AC & \multicolumn{1}{c|}{F1} & AC & F1 \\ \midrule
    \textbf{ET-BERT} & {\color[HTML]{CB0000} 42.0} & \multicolumn{1}{c|}{{\color[HTML]{CB0000} 38.9}} & 59.2 & 54.3 & {\color[HTML]{CB0000} 20.5} & \multicolumn{1}{c|}{{\color[HTML]{CB0000} 13.8}} & 55.3 & 51.5 \\
    \textbf{YaTC} & {\color[HTML]{CB0000} 25.5} & \multicolumn{1}{c|}{{\color[HTML]{CB0000} 25.1}} & \textbf{60.0} & \textbf{54.8} & {\color[HTML]{CB0000} 34.0} & \multicolumn{1}{c|}{{\color[HTML]{CB0000} 27.8}} & 77.3 & 74.8 \\
    \textbf{NetMamba} & {\color[HTML]{CB0000} 15.6} & \multicolumn{1}{c|}{{\color[HTML]{CB0000} 13.6}} & 52.4 & {\color[HTML]{CB0000} 48.6} & {\color[HTML]{CB0000} 16.9} & \multicolumn{1}{c|}{{\color[HTML]{CB0000} 11.3}} & 78.3 & 76.0 \\ 
    \textbf{TrafficFormer} & {\color[HTML]{CB0000} 39.2} & \multicolumn{1}{c|}{{\color[HTML]{CB0000} 36.9}} & 53.7 & {\color[HTML]{CB0000} 49.2} & {\color[HTML]{CB0000} 46.3} & \multicolumn{1}{c|}{{\color[HTML]{CB0000} 42.3}} & 71.4 & 69.2 \\ 
    \textbf{netFound} & {\color[HTML]{CB0000} 22.9} & \multicolumn{1}{c|}{{\color[HTML]{CB0000} 18.8}} & 56.6 & 52.4 & {\color[HTML]{CB0000} 28.0} & \multicolumn{1}{c|}{{\color[HTML]{CB0000} 22.9}} & \textbf{90.8} & \textbf{89.7} \\ \midrule
    \textbf{\tool} & \textbf{69.2} & \multicolumn{1}{c|}{\textbf{62.2}} & -- & -- & \textbf{71.3} & \multicolumn{1}{c|}{\textbf{68.1}} & -- & -- \\ \bottomrule
    \end{tabular}%
}
\caption{Flow-based classification tasks. Only per-flow split is possible. Similarly to the per-packet case, results improve only when the encoder is unfrozen.}
\label{tab:flow_classification}
\end{table}

When training the models, we follow the original papers' indications. For \textit{netFound}, we select the median burst for each flow and the median packet for each burst based on the distribution of the data sequence lengths, with a maximum of 12 bursts and 6 packets per burst. For the other models, we choose the first five packets of each flow. For \tool, being it a packet-level encoder, we adopt a simple majority vote scheme: Without additional training, we classify the first 5 packets of each flow and directly assign the flow class based on the majority of the labels of these 5 packets. We only use the encoder in a frozen manner. 

We adopt the same balanced split for training so that each class has a similar number of samples. This restricts the number of training samples per class to match the class with the fewest samples, stressing the models' few-shot learning capabilities.

We compare the results in Table~\ref{tab:flow_classification}. The same conclusions as in the previous experiment hold: First, all representation learning models struggle to classify the flow when the encoder is frozen. Once more, the representation learned during pre-training fails to effectively capture key features.
Performance improves when the encoder is unfrozen and allowed to update all its weights freely. However, even in this case, the results remain comparable to those of \tool, which achieves similar performance despite having a frozen encoder and relying on a simple majority-voting scheme. \textit{netFound} emerges as the best classifier in the TLS-120 task. This is due to its complex architecture (see next), and to specific pretext tasks that aim at including specific header fields (e.g., packet size) and flow-level multimodal information (e.g., time interval).

We noticed a large difference concerning the claimed results of \textit{ET-BERT} on \textit{TLS-120} (macro F1-Score of 97.5\%~\cite{lin2022etbert}).
This is due to: (i) the difference balanced strategy we adopt for training, which stresses the few-shot learning properties of the models; and (ii) the possible presence of the TLS Client Hello packet with plain text SNI (the authors do not confirm they removed it while in the public dataset it is not present) that would make the task simpler.
We ran an additional experiment with the original unbalanced 8:1:1 per-flow split, and the performance increased accordingly.

\begin{figure}[t]
    \centering
    \includegraphics[width=\columnwidth, trim={0 1.6cm 0 0}]{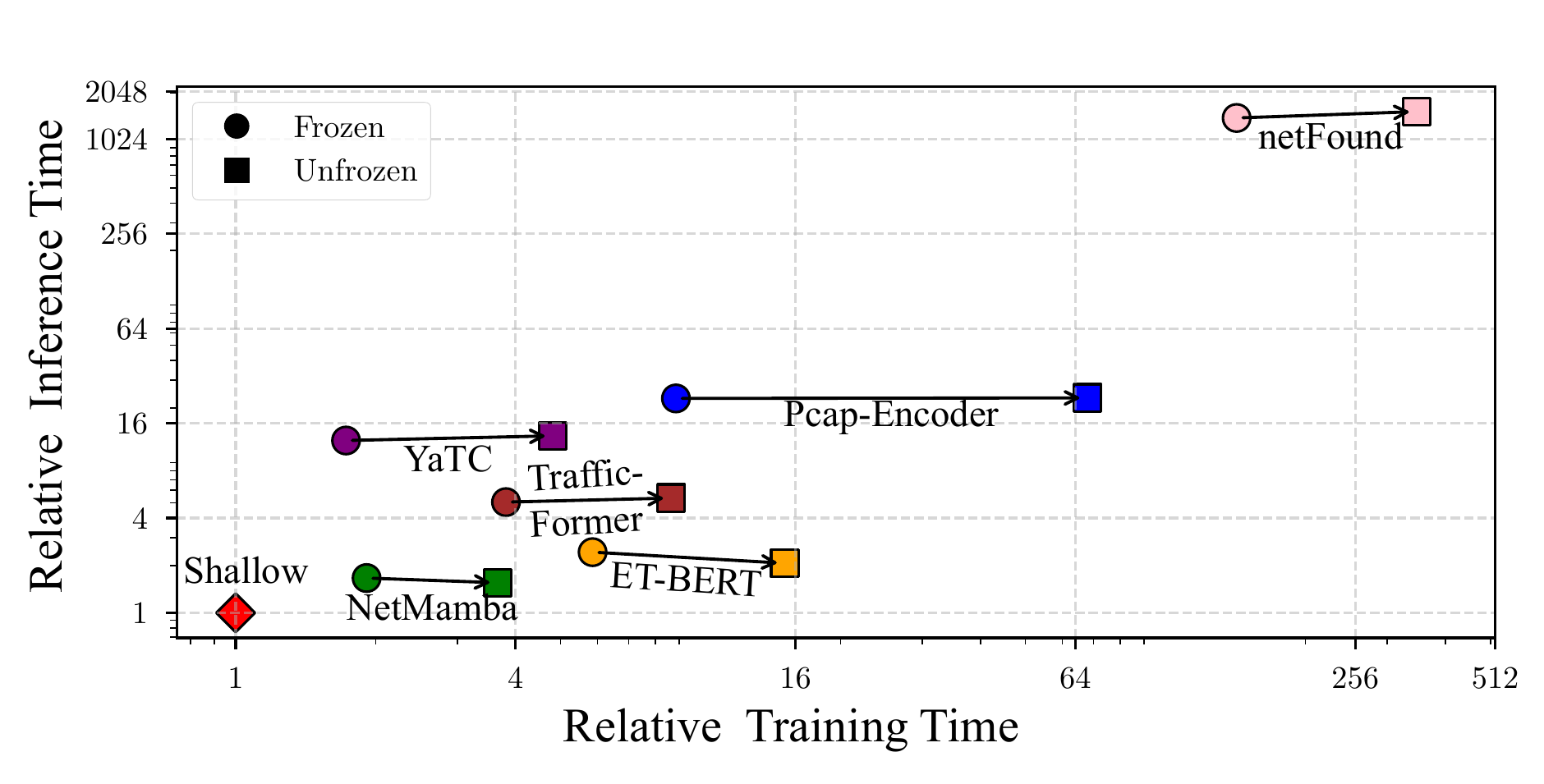}
    \caption{Relative training and inference times. All models are much slower than the shallow baseline, with larger models having the worst ratio (up to $2048\times$ slower at inference).}
    \label{fig:complexity}
\end{figure}

\subsection{Model efficiency}

We compare the inference and training time for the different models with frozen or unfrozen encoders. We measure the time to complete the 3-folds in the \textit{VPN-app}, per-flow split setup. 
Fig.~\ref{fig:complexity} reports the results normalized to the Random Forest which is the fastest \footnote{22 seconds for training and validation, and 5 seconds for testing.}.
At training time, all representation learning models require from 2 to 500 times more time than the RF. When comparing frozen versus unfrozen training, the time grows by a 2x - 8x factor, depending on the model.
At inference, the efficiency only depends on the complexity of the model. \textit{NetMamba} is the most efficient, while \textit{netFound} is the most costly model as it relies on the BERT Large architecture.

Overall, \tool good representation and classification performance is counterbalanced by it being the second slowest models, both at training and inference time. While the resource consumption of \tool is higher than that of shallow models, in scenarios where network data is becoming increasingly complex and protocols are becoming more diverse, this self-supervised model can extract information with stronger semantics, allowing it to better adapt to and generalize across different network environments and requirements.

%% file: sections/07_Related_Work/related_work.tex
Traffic classification has been a traditional problem since the Internet's birth. Initially solved by DPI~\cite{finsterbusch2013survey}, after the adoption of encryption, researchers started using machine learning~\cite{nguyen2008survey}, deep learning~\cite{rezaei_deep_2019,pacheco_towards_2019,papadogiannaki2021survey}, and recently representation learning to face it. 

Along the way, several works explicitly questioned some approaches and suggested best practices to follow. 
Regarding the problem of shortcut and `spurious correlations`, Arp et al.~\cite{arp2022and} studied top-tier security conference papers from the past decade, confirming that pitfalls are widespread and demonstrating how they can lead to unrealistic performance and interpretations. Similarly, Jacobs et al.~\cite{emperor_has_no_clothes} showed that models trained for Network Security can be shortcut learners, and propose a decision tree-based methodology to explain the model's predictions. More recently, Willinger et al.~\cite{willinger2025something} introduced the \textit{Credibility Crisis} currently impacting ML for networking, and discussed the \textit{underspecification} problem that affects standard supervised ML pipelines. Prior end-to-end works primarily address models trained with standard supervision; here we draw attention to what we call the \textit{sweet danger of sugar}. Self-supervision -- and specifically representation learning -- represents a methodological advancement over end-to-end supervised learning and it can therefore be a way-out to the credibility problem. However, as we show, poor application of self-supervised methods can lead to pitfalls similar to those found in supervised approaches.

Focusing on one cause of shortcut learning, \ie data issues, Dainotti et al.~\cite{dainotti2012issues} provided recommendations, including rigorous data collection and common benchmark definition. Authors do not explicitly mention the need for proper and careful data splitting. More recently, on the problem of spurious correlations, Flood et al.~\cite{flood2024bad} pinpointed issues in datasets commonly used for cybersecurity that can induce bias in intrusion detection systems. Similarly, Wickramasinghe et al.~\cite{SOK} identified severe limitations in commonly used datasets (\eg \textit{substantial portions of public datasets contain unencrypted traffic}). We are the first to identify how critical the dataset splitting is and the importance of focusing on frozen training.

Broadening the perspective, numerous studies have identified general pitfalls in machine learning, including shortcut learning~\cite{geirhos2020shortcut}, sampling bias and dataset shift~\cite{quinonero2022dataset,moreno2012unifying,tommasi2017deeper}, biased parameter selection~\cite{qin2017biased}, cherry picking~\cite{morse2010cherry}, and flawed training~\cite{McCoy2019RightFT} or evaluation practices~\cite{forman2010apples}. Our work builds on this foundation, offering recommendations for the specific challenges of representation learning in traffic classification and supporting the networking community in developing AI-based solutions and shared best practices.

%% file: sections/06_Conclusions/conclusions.tex
In this paper, we presented a detailed analysis of state-of-the-art representation learning works for traffic classification. 
We highlighted major pitfalls that previous works ignored, likely intoxicated by the `sugar' of the results falsely close to perfection.
After showing that the representation produced by the previous models is not informative, we advocate a frozen testing setup.
In fact, \tool is the only model that produces a useful representation for downstream traffic classification tasks. However, its complexity and performance, on par with shallow models, question its practicality for current problems.

Our findings offer relevant lessons for two communities: (i) networking researchers applying AI techniques, and (ii) AI researchers developing models for networking problems. Some insights—such as the effects of freezing strategies and dataset splitting—are broadly applicable and may generalize to adjacent fields like cybersecurity. Others, like distinguishing between packet- and flow-level analysis, removing extraneous protocols, and designing networking-specific pre-training tasks, are more domain-specific.

We hope that our work sheds light on the debunking and understanding of the correct usage of AI-based solutions in the context of traffic classification and the networking field.

\section*{Ethics} This work does not raise ethical issues.

%% file: sections/Appendix_tool.tex

Appendices are supporting material that has not been peer-reviewed.

\subsection{\tool details}
\label{sec:appendixEncoder}

\subsubsection{Packet representation learning task formalization}

We use the same problem formalization as the one in \cite{meng2022packet}. Consider a packet set $\mathcal{X} = \{\boldsymbol{x_1}, \boldsymbol{x_2},...,\boldsymbol{x_n}\}$ where $\boldsymbol{x_i}$ is the i-th packet, and each packet $\boldsymbol{x_i}$ is represented as an input sequence $\boldsymbol{x_i} = \{t_{i,1}, t_{i,2},..., t_{i,n}\}$ where $t_{i,j}$ is the j-th token derived from the split of the tokenizer. The length of the vector $\boldsymbol{x_i}$ can vary since the size of packets is unfixed.
In addition, each packet $\boldsymbol{x_i}$ can assume a number $n$ of labels $\boldsymbol{y_i}$ depending on the number of downstream tasks we want to solve. So, for a downstream classification task $j$, ${y}_{i,j} \in \mathcal{C}_j$.

To obtain the classification from the packet, we need the latent vector (packet representation) $\boldsymbol{r_i} \in \mathbb{R}^d$ where $d$ is the hidden dimension of the model. The latent vector is obtained by combining the columns of the latent matrix $\boldsymbol{H_i} \in \mathbb{R}^{d\times L}$ where $L$ is the number of tokens in each packet and $d$ the dimension of a single token. 

The formalization of the packet representation learning task becomes:
Given the input $\mathcal{X}$ and the corresponding label set $\mathcal{Y}$ of multiple classification tasks, the goal is to learn a single packet representation encoder $f:\boldsymbol{x_i} \rightarrow \boldsymbol{r_i}$ 
that obtain accurate $\boldsymbol{y_i}$ on downstream tasks by a function $g:\boldsymbol{r_i} \rightarrow \boldsymbol{y_i}$.

\subsubsection{Bottleneck}

T5 provides a single representation of size 768 for each of the $L$ tokens of the packet. Let's call $\boldsymbol{e_{i,j}}$ the representation of token $j$ for packet $i$. 
However, we want to obtain a single representation $\boldsymbol{r_i}$ of dimension 768 for the entire packet $\boldsymbol{x_i}$ from the $L$ representations of the tokens.  
Hence, we need to use an aggregator or bottleneck.
This dimensionality reduction process necessarily discards a part of the information set.
We tried different architectures:

\begin{enumerate}
    \item \textbf{First pooling}: a dummy solution that takes as packet representation the embedding of the first token, that is always the initial part of the question needed by T5. The representation vector of packet $i$ becomes:
    $$
        \boldsymbol{r_i} = \boldsymbol{e_{i,0}}
    $$
    \item \textbf{Mean pooling}: performs the average over the hidden vectors $\boldsymbol{e_{i,j}}$ of the hidden matrix. The representation vector of packet $i$ becomes:
    $$
        \boldsymbol{r_i} = \frac{\sum_{j=1}^L \boldsymbol{e_{i,j}}}{L}
    $$
    \item \textbf{Luong attention} \cite{LuongAttention}: performs a weighted average of the embeddings. The weights, computed for each $e_{i,j}$, must be positive and the sum is 1. The representation vector of one packet becomes:
    $$
        w_{j} = \frac{\exp{(\boldsymbol{e_{j}}^\top\,\boldsymbol{q}})}{\sum_{z=1}^L \exp{(\boldsymbol{e_{z}}^\top \,\boldsymbol{q}})}\qquad\qquad \boldsymbol{r_i} = \sum_{j=1}^L w_{j} \boldsymbol{e_{j}}
    $$
    where $\boldsymbol{q}$ is a learnable query vector and $w_j$ is the weight associated with the embedding vector $\boldsymbol{e_j}$.
\end{enumerate}

The bottleneck is part of the trained model T5 (encoder+decoder). So, even if the bottleneck is very simple, the underlying layers can adjust their weights to create a meaningful representation. Therefore, having a computationally expensive bottleneck is redundant.

\subsubsection{Question answering dataset and results.}

We created a dataset with multiple tasks for the Q\&A phase starting from the datasets already described in Subsection~\ref{pcapencoder}. 

Table~\ref{tab:questionsQ&A} shows the 8 questions selected in the Q\&A dataset.
Some of the questions are retrieval tasks that need to find the answer in the context. Others consist of more complex tasks, such as the computation of the checksum, or the payload length. 

On this question dataset, we obtain an average accuracy of 98.2\% on the test, averaging over different tasks. 

\begin{table}[]
\centering
\footnotesize
\begin{tabular}{c|c}
\toprule
& \textbf{Questions on Packets}\\
\midrule
\textbf{Retrieval}& \makecell{\textit{Which is the TCP checksum?}\\ \textit{Which is the destination IPv4/IPv6 of the packet?}\\ \textit{Which is the source IPv4/IPv6 of the packet?}\\\textit{Which is the id of IPv4/IPv6?}\\\textit{Which is the time to live of IPv4/IPv6?}}\\\midrule
\textbf{Computational}& \makecell{\textit{Is the packet's IPv4/IPv6 checksum correct?}\\\textit{Which is the last byte of the header in the third layer?}\\\textit{Which is the length of the payload in the third layer?}}\\

\bottomrule
\end{tabular}%

\caption{Retrieval and computational questions for the Q\&A pre-training task.}
\label{tab:questionsQ&A}
\end{table}

\subsubsection{Ablation study on \tool pre-training.} 
Our ablation study examined the impact of different components in \tool pipeline across the two tasks VPN-App (16) and TLS (120). Table~\ref{tab:effectiveness_pretrain} shows a clear performance hierarchy starting from the highest with the complete model. Removing the autoencoder component led to a moderate decrease in performance, particularly noticeable in the TLS task with an accuracy drop of $\approx8\%$. Most strikingly, using only the base T5 model without any pre-training resulted in severely degraded performance, especially on the more complex TLS dataset where the accuracy plummeted to 8.5. These results strongly suggest that both pre-training components contribute meaningfully to the model's effectiveness, with the Q\&A component appearing to be particularly crucial for maintaining strong performance

\begin{table}[]
\resizebox{.7\columnwidth}{!}{%
\centering
\begin{tabular}{c|cc|cc}
\toprule
 & \multicolumn{2}{c|}{\textbf{VPN-app (16)}} & \multicolumn{2}{c}{\textbf{TLS (120)}} \\ \cmidrule{2-5} 
\multirow{-2}{*}{\textbf{\begin{tabular}[c]{@{}c@{}}Model\\ (Flow Split)\\\,\end{tabular}}} & \multicolumn{1}{|c}{AC} & F1 & \multicolumn{1}{|c}{AC} & F1 \\ \midrule
\textbf{Autoencoder + Q\&A} & 83.5 & \multicolumn{1}{c|}{71.0} & 71.0 & 63.7  \\ 
\textbf{Q\&A only} & 82.6 & \multicolumn{1}{c|}{72.1} & 63.6  & 57.2    \\ 
\textbf{T5-base} & 54.5 & \multicolumn{1}{c|}{{\color[HTML]{CB0000} 39.8}} & \color[HTML]{CB0000} 8.5  & \color[HTML]{CB0000} 2.5      \\ \bottomrule
\end{tabular}%
}
\caption{Results on the per-Flow Split scenario by freezing and removing the pre-training phases of \tool.}
\label{tab:effectiveness_pretrain}
\end{table}

\subsection{Models Hyperparameters}\label{sec:appendixHyperparameters}
In the following, we report the chosen hyperparameters for the seven models under analysis:

$\bullet$ \textit{ET-BERT}: We remove the Ethernet and IP header and TCP ports. We set the learning rate to $2\cdot10^{-5}$ with 20 epochs for fine-tuning the unfrozen model, and $2\cdot10^{-3}$ for 60 epochs for the frozen model. 

$\bullet$ \textit{YaTC}: We anonymize IP addresses and ports, and group the first five packets to construct the input matrix by padding or truncating packets if necessary, as in the original paper. We set the learning rate at $2\cdot10^{-3}$ and the batch size to 64 over 200 epochs, both for the frozen and unfrozen tests.

$\bullet$ \textit{NetMamba}: We use the same learning rate and data processed as \textit{YaTC}.

$\bullet$ \textit{TrafficFormer}: We randomize the IP address and TCP ports and follow the same training-set augmentation proposed in the paper. We set the learning rate to $2\cdot10^{-5}$ with 20 epochs for fine-tuning the unfrozen model, and $1\cdot10^{-4}$ for 60 epochs for the frozen model. At the same time, we set an early stop of 5 epochs, that is, if the model performance does not improve in 5 consecutive epochs, the model will be terminated early.

$\bullet$ \textit{netFound}: We generate tokens and extract flow metadata as in the original paper. We set the learning rate to $2.5\cdot10^{-6}$ for a single GPU or $1\cdot10^{-5}$ for four GPUs and fine-tune the unfrozen and frozen models for 100 epochs with early stopping for 6 epochs. 

$\bullet$ \textit{\tool}: We load the T5-base model with pre-trained weights.
For the encoder adaptation, we use the AdamW optimizer, a learning rate of $5\cdot10^{-4}$, a linear rate scaling, a batch size of 8, and we train for 15 epochs. For question-answering tasks, we keep the same learning rate and scaling. We train for 20 epochs with a batch size of 24.

$\bullet$ \textit{Shallow model}: We use 4 different ML models (\ie Random Forest, XGBoost, MLP, LightGBM) and we use AutoGloun~\cite{agtabular} to automatically select the best hyperparameters. As features, we construct a vector for each packet by extracting the values of selected protocol fields (padding missing fields).

\begin{table}
\centering
\footnotesize
\begin{tabular}{ll}
\toprule
\textbf{Protocol} & \textbf{Packet fields} \\
\midrule
\textbf{IPv4} & \makecell{\textit{Source and Destination addresses, Type of service,}\\ \textit{Internet Header Length, ID, Checksum, Flags,}\\ \textit{Length, Protocol, Version, TTL, Fragmentation}} \\
\textbf{IPv6} & \makecell{\textit{Source and Destination addresses, Flow label,}\\ \textit{Version, Payload Length, Hop Limit,}\\ \textit{Traffic Class, Next Header}} \\
\textbf{UDP} & \makecell{\textit{Source and Destination ports, Checksum, Length}} \\
\textbf{TCP} & \makecell{\textit{Source and Destination ports, Timestamp,}\\ \textit{Window, Urgent pointer, Data offset, Flags,}\\ \textit{Checksum, Sequence and Acknowledgment numbers, Options}} \\
\bottomrule
\end{tabular}
\caption{Packet fields selected for the Shallow model training. Features are extracted from raw traces using the Python Scapy package (\url{https://scapy.net}).}
\label{tab:featureshallow}
\end{table}

\begin{table*}
\centering
\footnotesize
\begin{tabular}{
>{\columncolor[HTML]{FFFFFF}}c |
>{\columncolor[HTML]{FFFFFF}}c |
>{\columncolor[HTML]{FFFFFF}}c |
>{\columncolor[HTML]{FFFFFF}}c |
>{\columncolor[HTML]{FFFFFF}}c }
\toprule
\textbf{Type} & \textbf{Protocols} & \textbf{ISCX-VPN} & \textbf{USTC-TFC} & \textbf{CSTN-TLS1.3} \\ \midrule
\textbf{\begin{tabular}[c]{@{}c@{}}link-local\\ protocols\end{tabular}} & llmnr, nbns, mdns, lsd & \begin{tabular}[c]{@{}c@{}}922347\\ (3.45\%)\end{tabular} & \begin{tabular}[c]{@{}c@{}}227413\\ (3.93\%)\end{tabular} & 0 \\ \hline
\textbf{\begin{tabular}[c]{@{}c@{}}network management \\ protocols\end{tabular}} & \begin{tabular}[c]{@{}c@{}}icmp, icmpv6, dhcp, dhcpv6,\\ igmp, snmp, arp, cops\end{tabular} & \begin{tabular}[c]{@{}c@{}}437020\\ (1.63\%)\end{tabular} & \begin{tabular}[c]{@{}c@{}}373641\\ (6.46\%)\end{tabular} & 0 \\ \hline
\textbf{\begin{tabular}[c]{@{}c@{}}nat \\ protocols\end{tabular}} & nat-pmp, rsip, stun & \begin{tabular}[c]{@{}c@{}}205660\\ (0.77\%)\end{tabular} & \begin{tabular}[c]{@{}c@{}}68\\ (\textless{}0.01\%)\end{tabular} & 0 \\ \hline
\textbf{\begin{tabular}[c]{@{}c@{}}route management \\ protocols\end{tabular}} & \begin{tabular}[c]{@{}c@{}}db-lsp, db-lsp-disc, pathport, stp,\\ bfd echo, bgp, ecmp, asap\end{tabular} & \begin{tabular}[c]{@{}c@{}}21914 \\ (0.08\%)\end{tabular} & \begin{tabular}[c]{@{}c@{}}656\\ (0.01\%)\end{tabular} & 0 \\ \hline
\textbf{\begin{tabular}[c]{@{}c@{}}service management \\ protocols\end{tabular}} & ssdp, lldp, srvloc, opa, cbsp & \begin{tabular}[c]{@{}c@{}}8696 \\ (0.03\%)\end{tabular} & \begin{tabular}[c]{@{}c@{}}3812\\ (0.07\%)\end{tabular} & 0 \\ \hline
\textbf{\begin{tabular}[c]{@{}c@{}}real time\\ protocols\end{tabular}} & rtcp & \begin{tabular}[c]{@{}c@{}}2763\\ (\textless{}0.01\%)\end{tabular} & 0 & 0 \\ \hline
\textbf{\begin{tabular}[c]{@{}c@{}}network time\\ protocols\end{tabular}} & ntp & \begin{tabular}[c]{@{}c@{}}2386\\ (\textless{}0.01\%)\end{tabular} & \begin{tabular}[c]{@{}c@{}}35\\ (\textless{}0.01\%)\end{tabular} & 0 \\ \hline
\textbf{\begin{tabular}[c]{@{}c@{}}link management\\ protocols\end{tabular}} & llc, ipxsap & \begin{tabular}[c]{@{}c@{}}1582 \\ (\textless{}0.01\%)\end{tabular} & 0 & 0 \\ \hline
\textbf{\begin{tabular}[c]{@{}c@{}}distributed\\ protocols\end{tabular}} & thrift, dcerpc, rmi & \begin{tabular}[c]{@{}c@{}}182\\ (\textless{}0.01\%)\end{tabular} & \begin{tabular}[c]{@{}c@{}}5\\ (\textless{}0.01\%)\end{tabular} & 0 \\ \hline
\textbf{\begin{tabular}[c]{@{}c@{}}security\\ protocols\end{tabular}} & ocsp, pkix-cert, egd, chargen, tpm, knet & \begin{tabular}[c]{@{}c@{}}170\\ (\textless{}0.01\%)\end{tabular} & \begin{tabular}[c]{@{}c@{}}134\\ (\textless{}0.01\%)\end{tabular} & 0 \\ \hline
\textbf{\begin{tabular}[c]{@{}c@{}}industrial\\ protocols\end{tabular}} & \begin{tabular}[c]{@{}c@{}}r-goose, dcp-pft, dcp-af, vicp\\ nxp 802154 sniffer, enip, c1222, ax4000\end{tabular} & \begin{tabular}[c]{@{}c@{}}107\\ (\textless{}0.01\%)\end{tabular} & \begin{tabular}[c]{@{}c@{}}61\\ (\textless{}0.01\%)\end{tabular} & 0 \\ \hline
\textbf{\begin{tabular}[c]{@{}c@{}}remote access\\ protocols\end{tabular}} & vnc, x11, msnms & \begin{tabular}[c]{@{}c@{}}75\\ (\textless{}0.01\%)\end{tabular} & \begin{tabular}[c]{@{}c@{}}6  \\ (\textless{}0.01\%)\end{tabular} & 0 \\ \hline
\textbf{\begin{tabular}[c]{@{}c@{}}file\\ protocols\end{tabular}} & \begin{tabular}[c]{@{}c@{}}lanman, bjnp, spoolss, ndps,\\ laplink, bzr, cvspserver\end{tabular} & \begin{tabular}[c]{@{}c@{}}62\\ (\textless{}0.01\%)\end{tabular} & \begin{tabular}[c]{@{}c@{}}129\\ (\textless{}0.01\%)\end{tabular} & 0 \\ \hline
\textbf{\begin{tabular}[c]{@{}c@{}}quake\\ protocols\end{tabular}} & \begin{tabular}[c]{@{}c@{}}quake, quake2, quake3, \\ quakeworld\end{tabular} & 0 & \begin{tabular}[c]{@{}c@{}}4\\ (\textless{}0.01\%)\end{tabular} & 0 \\ \hline
\textbf{\begin{tabular}[c]{@{}c@{}}mobile\\ protocols\end{tabular}} & gsm, ipa, gtp & 0 & \begin{tabular}[c]{@{}c@{}}18\\ (\textless{}0.01\%)\end{tabular} & 0 \\ \hline
\textbf{\begin{tabular}[c]{@{}c@{}}iot management\\ protocols\end{tabular}} & \begin{tabular}[c]{@{}c@{}}bat.vis, tplink-smarthome,\\ coap,mqtt\end{tabular} & 0 & \begin{tabular}[c]{@{}c@{}}11\\ (\textless{}0.01\%)\end{tabular} & \begin{tabular}[c]{@{}c@{}}10\\ (\textless{}0.01\%)\end{tabular} \\ \hline
\textbf{\begin{tabular}[c]{@{}c@{}}others\\ protocols\end{tabular}} & tds, bitcoin & 0 & \begin{tabular}[c]{@{}c@{}}7\\ (\textless{}0.01\%)\end{tabular} & 0 \\ \bottomrule
\end{tabular}
\caption{The protocols we filter and number and percentage (in parenthesis) of removed packets for each dataset.}
\label{tab:protocols_filter}
\end{table*}

\subsection{Shallow models details}\label{shallow_models_detail}
Table~\ref{tab:featureshallow} lists the fields selected for the training of shallow models. The specific fields vary across datasets based on the presence or absence of the different datalink and transport protocols.

\subsection{Filter Details}\label{filters}
In Table~\ref{tab:protocols_filter} we list all protocols we filter using Tshark filters. We detail the total number of packets each filter removes from each trace. The main protocols involved are closely related to network management, link-local communication (link-local), and NAT (especially STUN). The CSTN trace was already cleaned, while the other contains from 5 to 10\% of unrelated protocols.